# Solution-processed GaSe nanoflake-based films for photoelectrochemical water splitting and photoelectrochemical-type photodetectors

*Marilena Isabella Zappia,† Gabriele Bianca,† Sebastiano Bellani,†\* Michele Serri, Leyla Najafi, Reinier Oropesa-Nuñez, Beatriz Martín-García, Daniel Bouša, David Sedmidubský, Vittorio Pellegrini, Zdeněk Sofer, Anna Cupolillo and Francesco Bonaccorso\**

Marilena Isabella Zappia, Dr. Reinier Oropesa-Nuñez, Dr. Vittorio Pellegrini, Dr. Francesco Bonaccorso
BeDimensional Spa., via Albisola 121, 16163 Genova, Italy

Gabriele Bianca, Dr. Sebastiano Bellani, Dr. Michele Serri, Dr. Leyla Najafi, Dr. Beatriz Martín-García, Dr. Vittorio Pellegrini, Dr. Francesco Bonaccorso
Graphene Labs, Istituto Italiano di Tecnologia, via Morego 30, 16163, Genova, Italy
E-mail: sebastiano.bellani@iit.it; francesco.bonaccorso@iit.it

Gabriele Bianca
Dipartimento di Chimica e Chimica Industriale, Università degli Studi di Genova, via Dodecaneso 31, 16146 Genoa, Italy.

Dr. Daniel Bouša, Prof. David Sedmidubský, Dr. Zdeněk Sofer
Department of Inorganic Chemistry, University of Chemistry and Technology Prague, Technická 5, 166 28 Prague 6, Czech Republic

Dr. Anna Cupolillo, Marilena Isabella Zappia
Department of Physics, University of Calabria, Via P. Bucci cubo 31/C 87036, Rende (CS), Italy

†These authors equally contributed

**Abstract**
Gallium selenide (GaSe) is a layered compound, which has been exploited in nonlinear optical applications and photodetectors due to its anisotropic structure and pseudo-direct optical gap. Theoretical studies predicted that its two-dimensional (2D) form is a potential photocatalyst for water splitting reactions. Herein, we first report the photoelectrochemical (PEC) characterization of GaSe nanoflakes (single-/few-layer flakes), produced *via* liquid phase exfoliation, for hydrogen evolution reaction (HER) and oxygen evolution reaction (OER) in both acidic and alkaline media. In 0.5 M $H_2SO_4$, the GaSe photoelectrodes display the best PEC performance, *i.e.* a ratiometric power-saved metric for HER ($\Phi_{saved,HER}$) of 0.09% and a ratiometric power-saved metric for OER ($\Phi_{saved,OER}$) of 0.25%. When used as PEC-type photodetectors, GaSe photoelectrodes show a responsivity of ~0.16 A W$^{-1}$ upon 455 nm illumination at light intensity of 63.5 µW cm$^{-2}$ and applied potential of –0.3 V *vs.* reversible hydrogen electrode (RHE). The stability analysis of the GaSe photodetectors evidences a durable operation over tens of cathodic linear sweep voltammetry scans in 0.5 M $H_2SO_4$ for HER. *Viceversa*, degradation effects have been observed in both alkaline and anodic operation due to highly oxidizing environment and $O_2$-induced (photo-)oxidation effects. Our results provide new insight into PEC properties of GaSe nanoflakes for their exploitation in photoelectrocatalysis, PEC-type photodetectors and (bio)sensors.

# 1. Introduction

Gallium selenide (GaSe) is a layered pseudo-direct optical bandgap (direct transition positioned just above the indirect one) binary chalcogenide (namely, a group-III monochalcogenide)[1],[2] composed of vertically stacked Se–Ga–Ga–Se tetralayers held together by van der Waals forces.[3] Several stacking of tetralayered blocks (*i.e.*, GaSe monolayers) are possible and form various polytypes (β, γ, δ and ε) of the crystal.[4] The most common one is the hexagonal ε-GaSe (space symmetry group: $P\bar{6}m$-$D'_{3h}$),[5],[6],[7] which is grown by Bridgman methods.[8],[9] Due to its structure, GaSe shows fascinating optoelectronic properties,[10] including photoresponse in ultraviolet/visible (UV-vis) spectral range (from 1.8 to 5 eV),[10],[11],[12] non-linear optical behavior,[13],[13] and a distinctive spin physics (*e.g.*, spin-orbit coupling effects[14] and generation/retention of spin polarization under nonresonant optical pumping[15],[16]). For the aforementioned properties, GaSe has been proposed for photodetectors with high responsivity[17],[18],[19],[20],[21],[22],[23],[24],[25],[26] (*e.g.*, up to values > 1000 A W$^{-1}$ at light intensity ≤ 0.1 mW cm$^{-2}$, in heterojunction with graphene),[26] non-linear frequency generation (*e.g.*, second and third harmonic and ultra-broadband radiation generation),[27],[6],[28],[29],[30],[31],[32] spin polarization control (*e.g.*, spintronic logic devices),[33] light-emitting devices,[34],[35] optical microcavities[36] and saturable absorbers.[37],[38] Moreover, the number of layers and strain engineering strongly affect the GaSe optoelectronic properties,[39],[40],[41],[42] which can be on-demand tailored to fulfil the requirements of the final applications.[42],[43] In particular, theoretical calculations demonstrated a c-axis confinement-induced bandgap ($E_g$) blue shift, which resemble the behavior of transition metal dichalcogenides.[44],[45],[46],[47],[48] An indirect $E_g$ superior to 3 eV has been theoretically predicted for GaSe monolayer,[44],[45],[46],[47],[48] and confirmed by experimental measurements.[34],[49] Such values of $E_g$ raise interest for application of two-dimensional GaSe, as well as other group-III monochalcogenides, such as InSe[50], GeS[51], GeSe[52],[51] and GeTe[53], as photo(electro)catalysts for water splitting reactions.[42],[54] Actually, GaSe monolayer fulfils the fundamental requirements for a water splitting photo(electro)catalysts, *i.e.*: 1) conduction band minimum (CBM) energy ($E_{CBM}$) > reduction potential of $H^+/H_2$ ($E(H^+/H_2)$); valence band maximum (VBM) energy ($E_{VBM}$) < reduction potential of $O_2/H_2O$ ($E(O_2/H_2O)$).[42],[54] Moreover, the two-dimensional nature of GaSe flakes maximizes the surface area available for water splitting reactions.[42],[55],[56] Meanwhile, the distance between the photogenerated charges and the surface area is virtually reduced to zero,[42],[55],[56] suppressing electron-hole recombination losses.[57] The validation of the GaSe nanoflakes for PEC reactions could pave the way towards the design of novel GaSe-based PEC-type photodetectors/PEC sensors that can operate with low voltage sources, or even without external energy supply systems,[58],[59],[50],[60] as well as eliminate complex device manufacturing. Notably, by detecting an analyte *via* photo-induced electrochemical reactions, PEC sensors offer several advantages over electrochemical sensor, since they can operate in differential mode to reduce the background signal (down to the limit of lock-in detection noise)[60],[61] and avoid frequent recalibrations.[60],[62],[63],[64] Therefore, the validation of GaSe and other group-III monochalcogenides as novel PEC-active two-dimensional materials represents a great potential for the design of efficient photocatalysts and innovative optoelectronic devices.[65],[66],[67],[68],[69] Despite these encouraging driving factors, the PEC properties of GaSe are still experimentally uncharted. A certain reluctance to study the (photo)electrochemical properties of GaSe and other group-III monochalcogenides undoubtedly originated from their tendency to undergo surface oxidation.[41],[70],[71],[72],[73],[74],[75] The latter can occur either in a two-step reaction: $GaSe + 1/4O_2 = 1/3Ga_2Se_3 + 1/6Ga_2O_3$ followed by $Ga_2Se_3 + 3/2O_2 = Ga_2O_3 + 3Se$;[76] or in a single-step reaction: (2) $GaSe + 3/4O_2 = 1/2Ga_2O_3 + Se$.[77] The presence of humidity and light has been reported to accelerate further the surface oxidation of GaSe.[71],[72] However, it is worth noticing that the material oxidation process can be controlled by properly adjusting (photo)electrochemical conditions such as potential, pH and the dissolved $O_2$ concentration. For example, electrochemical reduction has been exploited to decrease the surface oxidation of GaSe crystals.[78] Moreover, $Ga_2O_3$ can dissolve to $Ga^{3+}$ and $GaO^{2-}$ or $GaO_3^{3-}$ in acidic and alkaline media, respectively,[79] restoring the chalcogenide phase at the surface.[78] Noteworthy, theoretical simulations predicted that the water solubility of GaSe is below $10^{-18}$ mol mL$^{-1}$,[42] which implies that GaSe can be stable in water.

Driven by the aforementioned considerations, we aimed to unveil the photoelectrochemical behaviour of GaSe nanoflakes. We theoretically investigated the electronic structure of GaSe nanoflakes by using density functional theory (DFT) calculations. The calculated $E_g$ values indicate that GaSe nanoflakes can absorb a significant fraction of solar irradiation. By determining the $E_{CBM}$ and the $E_{VBM}$ of GaSe nanoflakes and their alignment with the redox potential of $H^+/H_2$ and $O_2/H_2O$, respectively, we predicted that GaSe nanoflakes should act as photocatalysts for water splitting reactions. To prove our theoretical predictions, GaSe nanoflakes were prepared in liquid dispersion by means of a viable and environmentally friendly liquid-phase exfoliation (LPE)[80],[81] of synthesized bulky crystals in anhydrous 2-Propanol (IPA). Due to its scalability, the LPE method is suitable for applications in which large amounts of material are needed, overcoming the low-throughput suffered by micromechanical cleavage-based exfoliation[82] and bottom-up nanomaterial synthesis (*e.g.*, CVD),[83],[84],[85] which are typically exploited for fundamental studies, including those on photodetectors.[25],[44],[73],[86], The as-produced GaSe nanoflakes were exploited as solution-processable materials for PEC water splitting and PEC-type photodetectors in aqueous electrolytes. In particular, the photoelectrodes were fabricated by spray coating the GaSe nanoflakes dispersion onto graphite paper. Our device fabrication does not involve any nanofabrication step, as typically devised to design Ga-based thin-film transistor for optoelectronic applications.[25],[44],[86],[87] The as-produced photoelectrodes were investigated for PEC HER and OER in 0.5 M $H_2SO_4$ (pH 1) and 1 M KOH (pH 14) under the simulated sunlight (*i.e.*, AM 1.5G illumination). The PEC properties of the GaSe nanoflakes were exploited to conceive PEC-type photodetectors, investigating the response at fixed illumination wavelength (455, 505 and 625 nm) in both acidic and alkaline media.

## 2. Result and discussion

### 2.1. Understanding the PEC properties of GaSe nanoflakes for water splitting reactions

The thermodynamic requirements for a water splitting photo(electro)catalyst are $E_{CBM} > E(H^+/H_2)$ and $E_{VBM} < E(O_2/H_2O)$ for HER and OER, respectively.[88],[89] In order to evaluate if GaSe crystals and nanoflakes fulfil these requirements, we performed electronic structure calculations using DFT with generalized gradient approximation (GGA-PBE96)[90] and Heyd-Scuseria-Ernzerhof hybrid exchange-correlation functional (HSE06)[91] for bulk and 1-, 2-, 4- and 6-layer (denoted as B and 1L, 2L,..., 6L)-GaSe (see details in Supporting Information, Experimental section). As we will show below, the electronic structure of 6L-GaSe is similar to the one of bulk GaSe (B-GaSe). Therefore, xL-GaSe with x>6 were not investigated by DFT calculation, since they display electronic properties resembling those of the B-GaSe. Our calculations show that B-GaSe is a direct bandgap semiconductor with VBM and CBM at the Γ-point of the first Brillouin zone. It is worth noticing that the sole use of GGA-PBE96 to describe exchange-correlation interaction between the electrons underestimates the material $E_g$ (around ~1 eV), since it neglects the screened Coulomb potential for Hartree–Fock exchange.[92],[93] By contrast, the use of hybrid HSE06 functionals (**Figure 1a**) results in an $E_g$ of 1.91 eV, which is similar to experimental values reported in literature (between 1.9 and 2.0 eV).[11],[94],[95] By reducing the number of layers in the GaSe, the $E_g$ progressively increases up to the highest value in 1L-GaSe (see Supporting Information, **Figure S1**), *i.e.*, 2.19 eV with the GGA-PBE96 and 3.14 eV with HSE06 (**Figure 1b**). Notably, this trend is accompanied by a shift of the VBM outwards from the Γ-point, leading to a direct-to-indirect bandgap transition in single-/few-layer GaSe (generically denoted with xL-GaSe, in which 1 ≤ x ≤ 6). Both B-GaSe and xL-GaSe exhibit a broad VB (energy ranging from 0 to –6 eV) formed by Se-4$p$ states, while their CB is given by both Ga-4$p$ and 4$s$ states. The valence band significantly contributes to the electron density localized around the lines connecting Ga and Se atoms (**Figure 1c**), a characteristic feature of covalent semiconductors. Due to the direct Ga-Ga bonding and the formal oxidation state of Ga (Ga(+II)), Ga-

4s states constitute a narrow band centred at –7.5 eV occupied by 2 electrons per unit cell comprising 2 Ga atoms. A 1.8 eV wide band of Se-4s character is located at ~12.5 eV below the Fermi level ($E_F$) and slightly above the sharp band constituted by highly localized Ga-3d states. The Ga-3d semi-core states substantially increases the VB electron density on Ga atoms (see the red spots on 2D section of electron density map of Ga-layer in **Figure 1d**) compared to Se layer (**Figure 1e**).

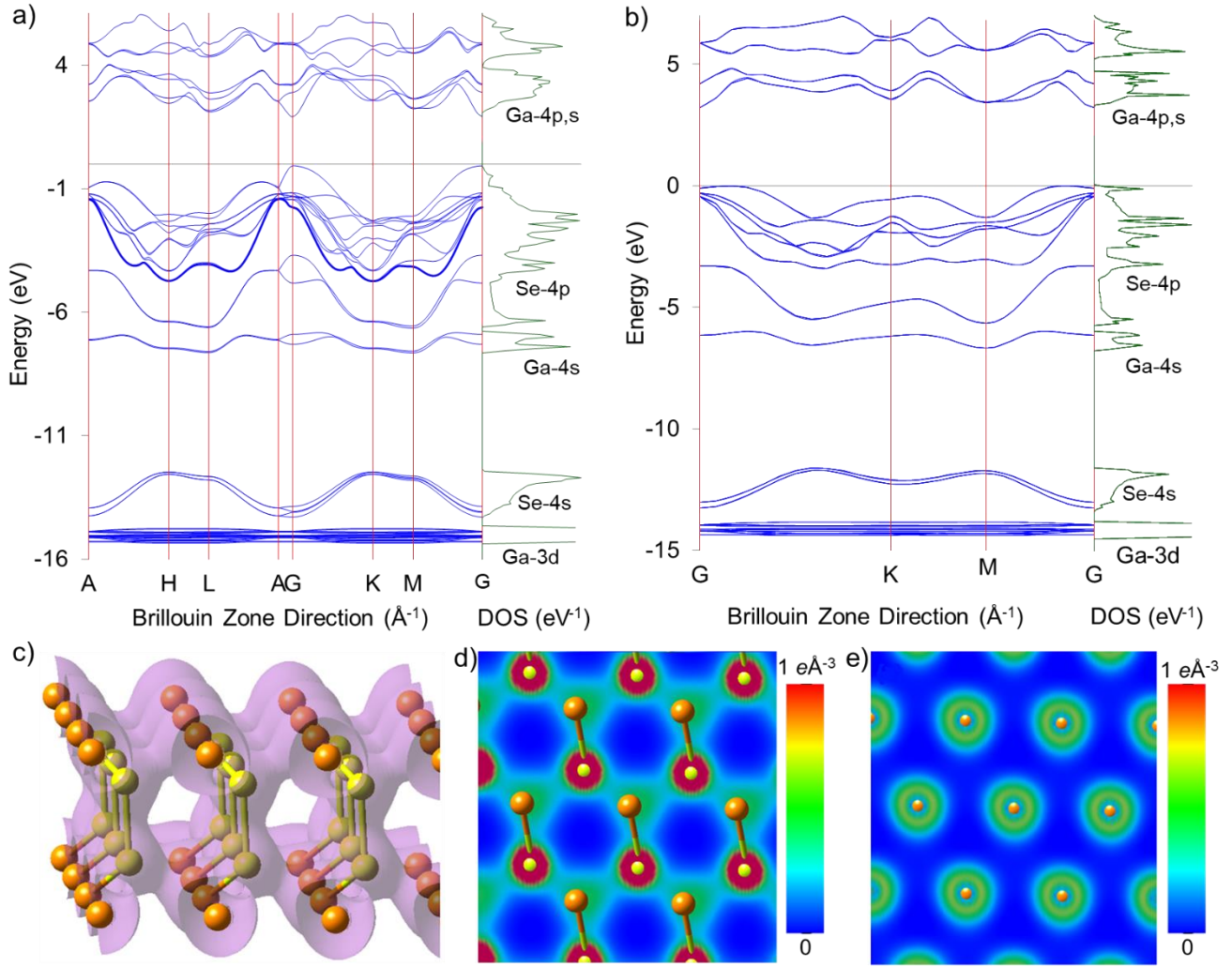

**Figure 1.** a,b) Band dispersion along the principal directions of the first Brillouin zone (blue lines) and the projected density of states (DOS) (green lunes) for B-GaSe 1L-GaSe, respectively, calculated by DFT using the HSE06 hybrid functionals. c) 3D iso-surface of the electron density = – 0.3 $e$ Å$^{-3}$. d,e) Electron density distributions in 2D cross section over Ga and Se layers, respectively.

The electron energies resulting from the calculation can be referred to the vacuum energy level in order to evaluate the $E_{CBM}$ and the $E_{VBM}$ related to (001) surface of the B- and few-layer GaSe structures relatively to $E(H^+/H_2)$ and $E(O_2/H_2O)$, respectively. **Figure 2** shows the $E_{CBM}$ and the $E_{VBM}$ as functions of the number of layers including the bulk limits. The $E(H^+/H_2)$ and $E(O_2/H_2O)$ as functions of the pH are also shown. Noteworthy, xL-GaSe with x ≤ 4 are predicted to be pH-universal photocatalysts for HER, since they fulfil the $E_{CBM} > E(H^+/H_2)$ requirement independently by the pH. Meanwhile, xL-GaSe with x ≥ 4 are predicted to be pH-universal photocatalysts for OER, since they fulfil the $E_{VBM} < E(O_2/H_2O)$ requirement independently by the pH. These results indicate that mixed nanoflakes with different number of layers should act as "local" tandem water splitting systems. As representative case, B-GaSe and 3L-GaSe can attain the overall water splitting (*i.e.*, both HER and OER) at pH < 1.5 and pH > 7, respectively. Differently, overall water splitting is impossible for both

2L- and 1L-GaSe, since their $E_{VBM}$ is higher than $E(O_2/H_2O)$ (*i.e.*, they cannot carry out OER). However, they can effectively operate in tandem configurations with other xL-GaSe, *i.e.*, they can collect electron from the CB of x-GaSe with $E_{CBM} > E_{VBM}$, meanwhile transferring their photoexcited electrons toward the electrolyte by performing HER.

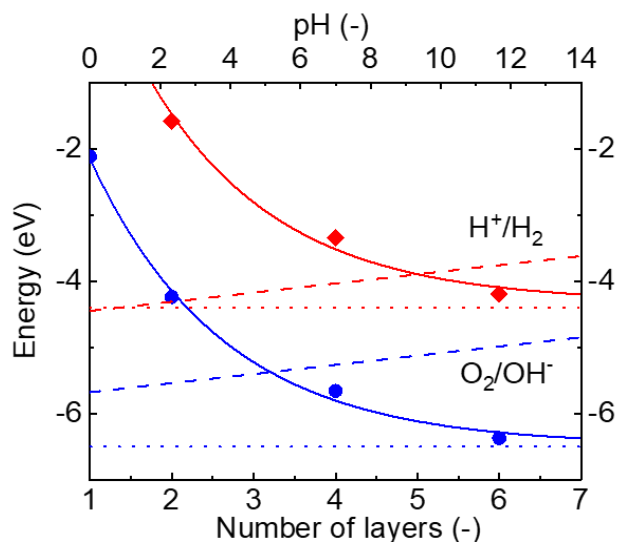

**Figure 2.** $E_{VBM}$ (Blue curve — and ● symbols) and $E_{CBM}$ (red curve — and ♦ symbols) of GaSe as a function of its layer number (bulk limits is also denoted by dotted lines), compared with the potentials of water splitting (*i.e.*, $E(H^+/H_2)$ and $E(O_2/H_2O)$) as a function of pH.

The van der Waals interactions, though weaker than covalent bonding inside the layers by more than three order of magnitudes, represent an essential feature of the layer cohesion in GaSe. While our DFT calculations using the bare GGA-PBE96 indicate that xL-GaSe is more stable than B-GaSe, the DFT-3D method, including the van der Waals dispersion correction,[96] estimates a surface energy of xL-GaSe more negative than the one of B-GaSe (by 10.5 kJ mol$^{-1}$ for 1L-GaSe), with positive surface energies of the single/few-layer slabs (*e.g.*, 143 and 145 mJ m$^{-2}$ for 1L-GaSe and 6L-GaSe, respectively). The slab is the model representing the flake in the calculations, see Methods for details. Being twice the surface energy of the slabs,[97] their interlayer cleavage energy of few-layer slabs is ~0.29 J m$^{-2}$, which is lower than the one measured for graphite (~0.37 J m$^{-2}$).[98] These data indicate that xL-GaSe should be produced by cleaving the 3D counterpart (B-GaSe), in agreement with pioneering experimental works[19],[99] and similarly to graphite.[82],[100]

**2.2. Synthesis and exfoliation of GaSe crystals**

The GaSe crystal was synthesized by direct reaction from Ga and Se elements.[78],[101] More in detail, granulated Ga and Se with an exact elemental stoichiometry of 1:1 were loaded in a quartz glass ampoule, subsequently evacuated, secured and heated at 970 °C (*i.e.*, melting temperature of GaSe)[102] for 1 h (heating rate = 5 °C min$^{-1}$). The synthesis products were then cooled down to room temperature (cooling rate = 1 °C min$^{-1}$), obtaining the GaSe crystal (**Figure 3a**). The as-produced GaSe crystal was pulverized to be characterized by scanning electron microscopy (SEM) coupled with energy-dispersive X-ray spectroscopy (EDS). The high-magnification SEM image of the GaSe edge (**Figure 3b**) shows the layered structure of the crystal. The SEM/EDS analysis (**Figure 3c**) reveals a slight Ga-enriched phases of the GaSe crystals (Ga-to-Se atomic ratio ~1.4, **Table S1**), which is in agreement with previous studies.[78],[101] The stoichiometric excess of Ga is attributed to the formation oxides (*i.e.*, $Ga_2O_3$), which partially passivate the GaSe surface, preventing the underlying GaSe from further oxidation.[70],[78],[79], The crystal structure of the GaSe crystals was evaluated by X-ray diffraction (XRD) measurements. The XRD pattern (**Figure S2**) agrees with the JCPDS 37-931 card, indicating that the as-synthesized GaSe crystals are in the form of the lowest energy polytype, *i.e.*, the hexagonal ε-GaSe[10],[103] (space symmetry group: $P\bar{6}m$-$D'_{3h}$),[5],[6],[7] in agreement with previous literature reporting similar GaSe crystal syntheses.[78],[101] The GaSe nanoflakes were produced by LPE[80],[81] of the synthesized crystals in anhydrous IPA followed by sedimentation-based separation (SBS)[104,105] to remove un-exfoliated crystals by ultracentrifugation (see Experimental section). Noteworthy, first principle calculations estimated a weak interlayer coupling in GaSe crystals (*i.e.*, cleavage energy ~0.33 J m$^{-2}$),[103] in agreement with our DFT calculations (cleavage energy of ~0.29 J m$^{-2}$ for xL-GaSe with x ≥2), indicating their feasible exfoliation similar to other layered materials, *e.g.*, graphite (cleavage energies of ~0.37 J m$^{-2}$, experimental value)[98] and MoS$_2$ (theoretical cleavage energy of 0.27 J m$^{-2}$).[97] The use of IPA as solvent has been reported to be effective for exfoliating another Ga-based group-III monochalcogenides (*i.e.*, GaS),[106] as well as other transition metal monochalcogenides (*e.g.*, InSe[107]) and dichalcogenides (*e.g.*, MoS$_2$,[108],[109],[110] MoSe$_2$,[111],[112] NbS$_2$[113]). Moreover, IPA is a low toxicity[114] and low boiling point (82.5 °C) solvent,[114] which allows the exfoliated material to be processed at low-temperature (< 100 °C) without special precautions (*e.g.*, use of fume hood and/or controlled atmosphere).[110] By starting from cost-effective artificial crystals, the LPE method does not recur to low-throughput micromechanical cleavage exfoliation[82] or time-consuming and expensive bottom-up synthesis (*e.g.*, CVD).[83],[84],[85] The morphology of the as-produced GaSe flakes was characterized by transmission electron microscopy (TEM) and atomic force microscopy (AFM) in order to evaluate their lateral dimension and thickness, respectively. **Figure 3d** shows the TEM image of representative GaSe nanoflakes, displaying wrinkled surfaces with irregular shapes, but edges with sharp profiles. **Figure 3e** shows an AFM image of a representative GaSe nanoflake, together with its height profile showing a step of ~3 nm at the edge. This height is attributed to a GaSe flake with less than 4 layers. In fact, the AFM thickness of a GaSe monolayer generally lies between 0.8 nm and 1 nm, depending on the substrate/GaSe interaction and the AFM instrumentation),[44],[115],[116],[49] and the GaSe interlayer distance is ~0.8 nm[4],[10],[47],[117]). Statistical TEM analysis (**Figure 3f**) indicates that lateral size data of the flakes follows a log-normal distribution peaking at ~45 nm, with maximum measured values above 0.5 µm. The statistical AFM analysis (**Figure 3g**) shows that the sample is mainly composed of single/few-layer GaSe nanoflakes. The estimated thickness values of the flakes are mainly between 1 nm (monolayer GaSe) and 10 nm, and follow a lognormal distribution peaked at ~2.4 nm. The crystal structure of the GaSe crystals and the exfoliated flakes was analyzed by Raman spectroscopy. As shown in **Figure 3h**, the GaSe crystals exhibit the two out-of-plane vibration modes $A^1_{1g}$ and $A^2_{1g}$ at ~134 and 308 cm$^{-1}$, respectively, and the two in-plane vibration mode $E^1_{2g}$ and $E^2_{1g}$ at ~212 and ~251 cm$^{-1}$, respectively, similar to previous reports.[43],[78],[118] In the case of the exfoliated GaSe flakes, the weakest $E^2_{1g}$ is not distinguishable from background signal, while $A^1_{1g}$ and $E^1_{2g}$ slightly shifts to lower and higher wavenumbers, respectively (see also statistical analysis in **Figure S3**). Theoretical studies show that the softening of the $A^1_{1g}$ and the strengthening of the $E^1_{2g}$ with the decrease of the thickness are related to the reduced inter-layer forces,[39] similarly to other transition metal chalcogenides (*e.g.*, MoS$_2$).[108] Moreover, our results agree with experimental works on the thickness dependence of the Raman spectrum of GaSe.[39],[43],[115] Notably, the $A^2_{1g}$ shows a red-shifts similarly to $A^1_{1g}$.[43] However, it is difficult to fully understand the behavior of this mode due to the presence of the second order mode of Si (*i.e.*, the substrate) at 302 cm$^{-1}$, as discussed in ref. [39]. Additionally, the Raman spectra of GaSe crystals and exfoliated flakes do not present signatures attributed to Ga$_2$Se$_3$, Ga$_2$O$_3$ and amorphous/crystalline Se (a-/c-Se) modes, which are observed at ~155,[119] ~200[120] and between 135–160 cm$^{-1}$,[121],[122],[123] respectively. This indicates that the LPE of GaSe crystal in anhydrous IPA does not cause significant additional surface oxidation of the native material.[73],[71],[77] This conclusion is further supported by X-ray photoelectron spectroscopy (XPS) analysis (**Figure S4,S5** and **Table S2-S7**). Since the Ga 2p photoelectrons have lower kinetic energy (K.E.) (~369 eV) and shorter inelastic mean free path (λ)

(~1.1 nm)[124,125] than the Ga 3d photoelectrons (K.E. ~1466 eV, λ ~3.0 nm), their different sampling depth allows to conclude that oxidation of GaSe is limited to outer layer of crystals/flakes. The capability to harvest solar light of the GaSe flakes was evaluated by performing diffusive reflectance spectroscopy (DRS) measurements. **Figure 3i** shows the diffusive reflectance (R) spectrum of the GaSe flakes film deposited on quartz substrate. The $E_g$ of the GaSe nanoflakes was determined using Kubelka-Munk theory of R phenomenon,[126],[127] *i.e.*, analysing the $(F(R)h\nu)^n$ *vs.* $h\nu$ (Tauc plot) (inset to **Figure 3i**) using the Tauc relation $(F(R)h\nu)^n = Y(h\nu - E_g)$, in which $F(R)$ is the Kubelka-Munk function (defined as $F(R) = (1-R)^2/2R$), $h$ is Planck's constant, $\nu$ is the photon's frequency, and $Y$ is a proportionality constant.[128] The value of the exponent denotes the nature of the electronic transition, discriminating between direct-allowed transition ($n = 2$) and indirect-allowed transition ($n = 0.5$).[109],[129] Due to the pseudo-direct gap behavior of GaSe,[1],[2] $n$ was set equal to 2. The estimated $E_g$ is 1.9 eV, as the one measured for GaSe crystal.[11],[94],[95] Ultraviolet photoelectron spectroscopy (UPS) measurements allowed to determine the Fermi level energy ($E_F$), *i.e.*, the WF, and the $E_{VBM}$.[109]

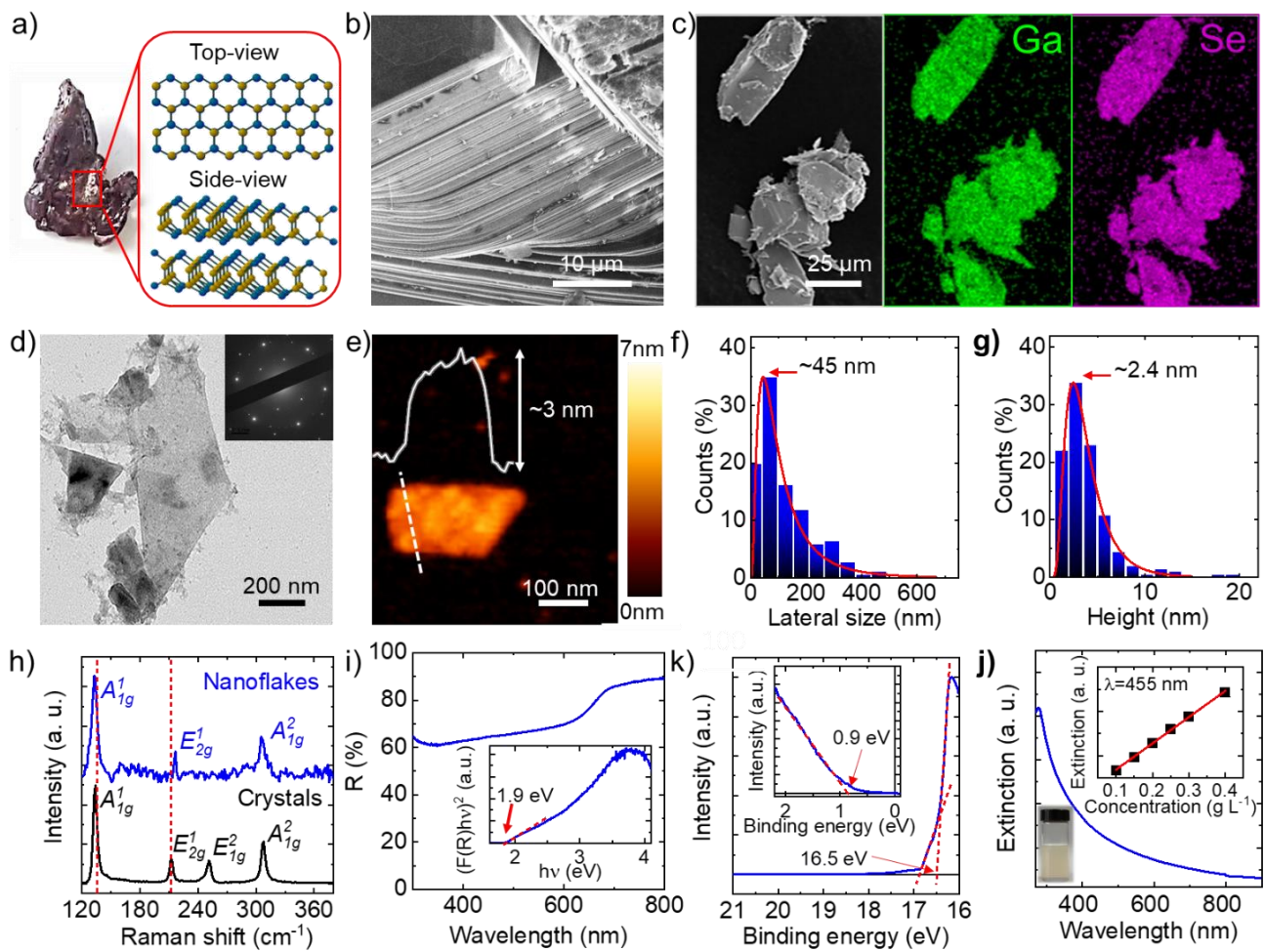

**Figure 3.** a) Photographs of the as-synthetized GaSe crystal. The crystal structure of the GaSe crystal (ε-GaSe) is also shown. b) SEM image of the GaSe crystals, evidencing the layered structure of its edge. c) SEM image of GaSe crystals and the corresponding EDS maps for Ga (green) and Se (violet). d) TEM image of the LPE-produced GaSe flakes. e) AFM image of a representative LPE-produced GaSe flake. The height profile of the indicated section (dashed line) is also shown. f) TEM statistical analysis of the lateral dimension of the exfoliated GaSe flakes. g) AFM statistical analysis of the thickness of the exfoliated GaSe flakes. h) Raman spectra of the as-synthetized GaSe crystals and the LPE-produced GaSe flakes. The Raman modes assigned to ε-GaSe are also shown. i) Spectrum of the diffusive reflectance (R) of the LPE-produced GaSe flakes. The inset shows the Tauc plot of the as-produced GaSe flakes. k) Secondary electron threshold region of He-I UPS spectrum of the LPE-produced GaSe flakes. The inset shows the He-I UPS spectrum region near the $E_F$ of the

GaSe flakes. j) Extinction spectrum (Ext(λ)) of the LPE-produced GaSe flakes dispersion. The top-right inset shows the Ext(λ) *vs.* c plot for λ = 455 nm of the GaSe flakes dispersions. The inset shows a photograph of the LPE-produced GaSe flakes dispersion.

**Figure 3k** shows that secondary electron cut-off (threshold) energies of the He I (21.22 eV) UPS spectrum is ∼16.5 eV, corresponding to WFs of 4.7 eV. The shoulder observed in the UPS spectrum can be attributed to the presence of surface oxides, *e.g.*, $Ga_2O_3$, which exhibits a n-type behaviour (corresponding to a WF < 4.5 eV) originated by oxygen vacancies.[130],[131],[132] The inset to **Figure 3k** shows the UPS spectra region near the $E_F$, which allows the $E_{VBM}$ to be estimated at ∼–5.6 eV. By considering the $E_g$ values estimated by the Tauc analysis, the $E_{CBM}$ is estimated at ∼–3.7 eV. It is worth noticing that the as-produced GaSe flakes have polydisperse morphology characteristics (see TEM and AFM analysis, **Figure 3d-e**). This means that the electronic properties of the thinnest flakes, as deduced by in the previous theoretical section, could be experimentally concealed by the thickest flakes,[133] which show the lowest $E_g$ and the highest $E_{VBM}$ (directly estimated by the Tauc and UPS analysis, respectively).[134] The concentration of the GaSe flakes dispersion was estimated by the Beer–Lambert law: $Ext(\lambda) = \varepsilon(\lambda)cL$, in which $Ext(\lambda)$ is the spectral extinction, $\varepsilon(\lambda)$ is the extinction coefficient, *c* is the material concentration and *L* is the optical path length.[135] More in detail, optical extinction measurements of controlled dilutions/concentrations of the as-produced GaSe flakes dispersion allow the extinction coefficient ($\varepsilon(\lambda)$) to be estimated form the slope of $Ext(\lambda)$ *vs.* c plot, being the slope = $\varepsilon(\lambda)L$.[106],[107] The concentration value of the as-produced dispersion (0.20 ± 0.02 g L$^{-1}$) was measured by weighting the solid material content in a known volume of the dispersion. **Figure 3j** reports the $Ext(\lambda)$ of the as produced GaSe flakes dispersion. The slope of the linear fitting of the $Ext(\lambda)$ *vs.* c plot (inset to **Figure 3j**) provides: $\varepsilon$(455 nm) = 113.0 L g$^{-1}$ m$^{-1}$. By using the experimentally derived $\varepsilon(\lambda)$ values, the concentration of GaSe flakes was fixed at 0.13 g L$^{-1}$ during the subsequent experiments. This value could underestimate the actual concentration of the solid material content (as measured by weight measurements), which also include by-products, *e.g.*, $Ga_2Se_3$, $Ga_2O_3$ and amorphous/crystalline Se (see Supporting Information, **Table S1** and **Figure S4**).

**2.3. Photoelectrochemical water splitting of GaSe photoelectrodes**

Based on the theoretical and experimental characterizations of their optoelectronic properties, the GaSe flakes were tested as candidate photoelectrocalysts for PEC water splitting and PEC-type photodetectors (**Figure 4a**). By taking advantage of the production of GaSe flakes in form of liquid dispersion in IPA, the GaSe-based photoelectrodes were fabricated by spray coating the GaSe flakes dispersion onto graphite paper (GaSe flakes mass loading = 0.1 mg cm$^{-2}$). **Figure 4b** shows a photograph of the as-produced GaSe photoelectrode, which was manually bent to show its flexibility. As shown by the SEM image (**Figure 4c**), the photoelectrode has a laminar structure, in which GaSe flakes preferentially orient horizontally to the terrace of the graphite paper.

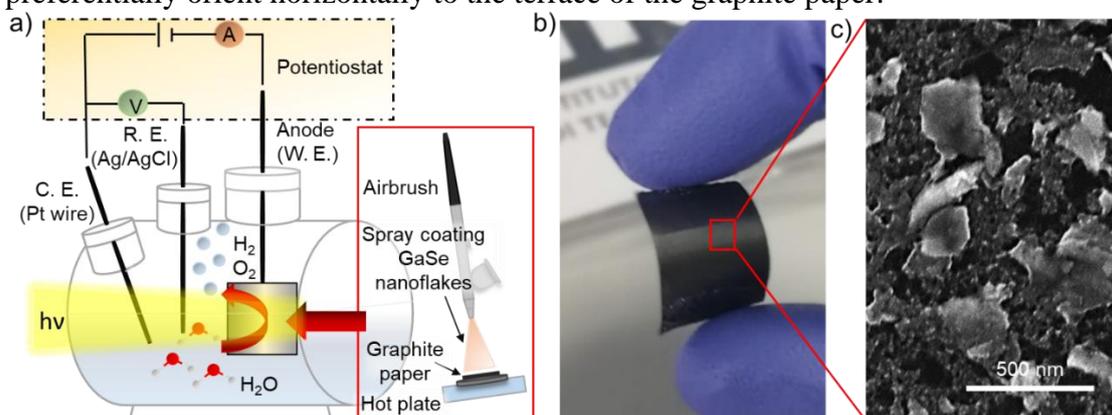

**Figure 4.** a) Schematic illustration of the GaSe photoelectrodes, produced by spray coating deposition of the LPE-produced GaSe flakes dispersion onto graphite paper (current collector), used for PEC

water splitting reaction and PEC-type photodetectors. b) Photograph of the GaSe photoelectrode and c) its corresponding SEM image.

The PEC water splitting activity of the as-produced photoelectrodes was evaluated in both acidic (0.5 M $H_2SO_4$) and alkaline (1 M KOH) $N_2$-purged solutions at room temperature. To the best of our knowledge, no previous study experimentally investigated the PEC properties of GaSe flakes in aqueous solutions, although they have been theoretically predicted by recent works.[42],[54] **Figure 5a,b** show the cathodic and anodic linear sweep voltammetry (LSV) scans for the as-produced GaSe photoelectrodes illuminated by chopped simulated sunlight (*i.e.*, AM 1.5G illumination) in 0.5 M $H_2SO_4$ and 1 M KOH. The common Figures of Merit (FoM) used to compare the performance of photoelectrodes are:[136] the positive onset potential ($V_{OP}$), the cathodic photocurrent density at 0 V *vs.* RHE ($J_{0V\ vs\ RHE}$), the anodic photocurrent density at +1.23 V *vs.* RHE ($J_{0V\ vs\ RHE}$), the ratiometric power-saved metric for HER ($\Phi_{saved,HER}$) and the ratiometric power-saved metric for OER ($\Phi_{saved,OER}$). The definition of these FoM is reported in the Supporting Information, Experimental section. In 0.5 M $H_2SO_4$, the photoelectrodes show a $V_{OP}$ of +0.14 V *vs.* RHE. The other FoM of the photoelectrodes are: $J_{0\ V\ vs\ RHE} = -9.3\ \mu A\ cm^{-2}$ and $\Phi_{saved,HER} = 0.09\%$ for HER; $J_{1.23\ V\ vs\ RHE} = +83.4\ \mu A\ cm^{-2}$ and $\Phi_{saved,OER} = 0.25\%$ for OER. In 1 M KOH, the photoelectrodes exhibit a clear photoanodic behavior, while the cathodic LSV scan shows a negligible photocurrent density and a significant negative dark current density ($< -10\ \mu A\ cm^{-2}$ for applied potential $< +0.2$ V *vs.* RHE). In our view, the alkaline media promote the formation of oxidized species, which can decompose in soluble products (*e.g.*, $GaO_2^-$) during cathodic LSV scans.[78],[79]

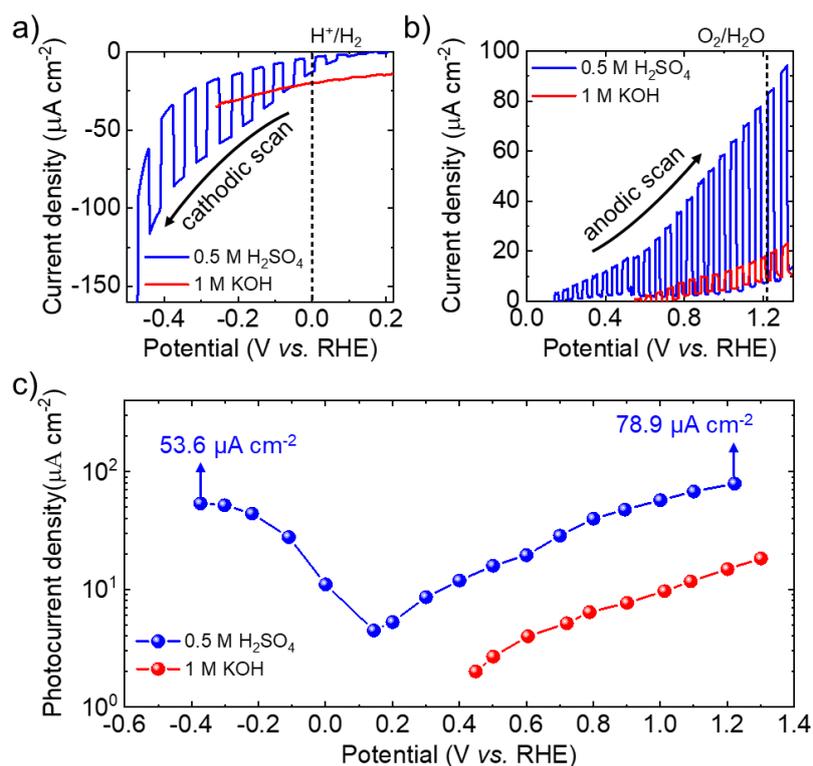

**Figure 5.** LSV curves measured for GaSe photoelectrodes for a) HER (cathodic scan) and b) OER (anodic scan) under chopped simulated sunlight (*i.e.*, AM 1.5G illumination) in 0.5 M $H_2SO_4$ and 1 M KOH. The redox potential for $H^+/H_2$ (0 V *vs.* RHE) and $O_2/H_2O$ (+1.23 V *vs.* RHE) are also indicated. c) Absolute photocurrent density of the photoelectrodes as a function of the applied potential measured in 0.5 M $H_2SO_4$ and 1 M KOH. The maximum absolute photocurrent density values obtained for both cathodic and anodic scan are also shown.

**Figure 5c** reports the absolute photocurrent density of the photoelectrodes as a function of the applied potential, showing maximum values in 0.5 M $H_2SO_4$, suggesting such medium for the subsequent development of PEC-type photodetectors. Although the above data indicate a convenient use of GaSe photoelectrodes for performing the PEC OER at potential significantly inferior to the redox potential of $O_2/H_2O$ (*i.e.*, 1.23 V *vs.* RHE), it is worth noticing that the evolved $O_2$ could degrade the photoelectrode surface by forming oxides.[41],[70],[71],[72],[73],[74],[75] Actually, it has been shown that the concomitant presence of $O_2$, humidity and above-gap illumination further accelerates the oxidation of GaSe flakes into $Ga_2O_3$, $SeO_2$ and a-Se *via* an intermediate stage of formation of gallium hydroxides (*i.e.*, $Ga(OH)_3$) and selenium oxide-water complexes and selenic acid.[72] In ref. [71], the water-assisted photo-oxidation was attributed to the transfer of photo-generated charge carriers towards aqueous $O_2$ (*i.e.*, dissolved $O_2$), generating highly reactive superoxide anions ($^\bullet O_2^-$) that rapidly degrade the optoelectronic properties of GaSe. However, by bubbling $O_2$ into the electrolytic solution, no oxygen reduction reaction (ORR) peak was observed during the LSV scan, which excludes an ORR-induced oxidation of the GaSe flakes. This is in agreement with the degradation mechanism proposed in ref. [72] (the PEC stability of our photoelectrodes will be further discussed here below for the case of PEC-type photodetectors).

**2.4. Photoelectrochemical-type photodetectors based on GaSe flakes**

On the basis of the preliminary results achieved for PEC water splitting, the GaSe photoelectrodes were investigated as PEC-type photodetectors in 0.5 M $H_2SO_4$ upon three different illuminations with wavelengths in the visible spectral range, *i.e.*, 455, 505 and 625 nm, whose energy is above the $E_g$ of the GaSe flakes (~1.9 eV (~653 nm), see Tauc analysis, **Figure 3i**). This means that the potential absorption from GaSe flakes can be explained by their optical transitions from valence to conduction bands (although optical transitions starting from intergap states, which are expected from the intrinsic p-type nature of GaSe,[21] cannot be excluded). Moreover, as a precaution, the applied potentials were limited between –0.3 V and +0.8 V *vs.* RHE to avoid highly reductive and oxidative condition possibly causing uncontrolled (photo)electrochemical degradation of the photoelectrodes. **Figure 6a** shows the responsivity of the GaSe photodetector as a function of the applied potential at each wavelength (light intensity = 63.5 µW cm$^{-2}$). As expected from the OAS analysis (**Figure 3i**), the responsivity of the GaSe photodetectors shows the maximum values (157 mA W$^{-1}$ at –0.3 V *vs.* RHE and 117 mA W$^{-1}$ at +0.8 V *vs.* RHE) under illumination at 455 nm, in agreement with absorption spectrum of the GaSe flakes in IPA dispersion. These values approach those of self-powered commercial UV-Vis photodetectors (*e.g.*, Si- or GaP-based photodiodes),[137] and are superior to those of relevant self-powered or low-voltage operating solution-processed photodetectors (**Table S2**).[50],[138],[139],[140],[141],[142],[143],[144]. The stability of the PEC response was evaluated by measuring the responsivity over repeated LSV scans. **Figure 6b** reports the responsivity retention measured at –0.3 V *vs.* RHE (cathodic regime) and +0.4 V *vs.* RHE (anodic regime). Clearly, the GaSe photodetector shows a durable responsivity (+35% after 20 LSV scans) under cathodic operation, while it progressively degrades during OER (–80% after 20 LSV scans). The photoelectrode degradation under anodic potentials is tentatively attributed to the progressive oxidation of the GaSe flakes caused by both anodic potential and evolved $O_2$.[78] Similar photodetector degradation under anodic potentials was observed in 1 M KOH (**Figure S6**), in which the devices have also shown inferior performance compared to the ones measured in 0.5 $H_2SO_4$. The Raman spectroscopy analysis of the electrodes before and after stability tests (**Figure S7**) further evidences that GaSe nanoflakes preserve their structural properties during cathodic operation in 0.5 M $H_2SO_4$. *Viceversa*, after anodic operation in acidic and alkaline media, the presence of peak attributed to $Ga_2O_3$ and the increase of intensity of the peak ascribed to elemental Se, respectively, support the degradation of GaSe nanoflakes, as indicated by the decrease of their OER-activity during successive LSV scans (see further details on Raman spectroscopy analysis in Supporting Information). Based on these results, the GaSe photodetectors were further characterized when working in cathodic regime and acidic media, in which they showed both optimal PEC-responses and electrochemical stability.

The photocurrent density of the GaSe photodetectors at fixed potential of −0.3 V *vs.* RHE was evaluated as function of the light intensity (**Figure 6c**). The photocurrent density increases with the light intensity passing from 0.011 to 31.8 mW cm$^{-2}$. The relationship between the photocurrent density and the light intensity is typically expressed by a power law, *i.e.*, photocurrent density ∝ (light intensity)$^{\gamma}$, in which γ is a factor determining the response of the photocurrent to light intensity.[145] For light intensity ≤ 56.7 μW cm$^{-2}$, the power equation fits to the experimental data with γ equal to 0.97, indicating an almost linear fit. Since a unity value for γ suggests negligible charge recombination and trapping processes,[145] it can be deduced that the two-dimensional morphology of GaSe flakes maximizes the surface area available for PEC reactions.[42],[55],[56] Meanwhile, the distance between the photogenerated charges and the material surface area, where electrochemical reaction occurs, is reduced close to zero (zero for the monolayer case),[42],[55],[56] suppressing electron-hole recombination losses.[57] Consequently, the responsivity (~0.16 A W$^{-1}$) is retained with increasing the light intensity (up to the tested values of 56.7 μW cm$^{-2}$). A representative condition of high light intensity (31.8 mW cm$^{-2}$) was also investigated, showing a decrease of the photodetector responsivity down to 19.5 mA W$^{-1}$. This effect could results from both trap-assisted and/or bimolecular charge recombination processes occurring at high light intensity regime,[145] and charge recombination effects in presence of kinetically limited heterogeneous charge transfer at solid/liquid interfaces.[50] The response time is another important FoM of photodetectors. The temporal response of our GaSe photoelectrodes was measured at −0.3 V vs. RHE (**Figure 6d**), showing rise time ($\tau_R$) and fall time ($\tau_F$) of 855 ms and 720 ms, respectively, which are between one and two order of magnitude larger than those of GaSe monolayer-based photodetectors (~20 ms),[19] but significantly inferior to solution processed PEC-type photodetectors (typically in the order of ten seconds).[50] Overall, our results suggest that GaSe flakes can be efficiently exploited for PEC applications, including water splitting and photodetectors, although stability issues have to be carefully taken into account to limit degradation effects in presence of $O_2$, water and above-gap illumination.

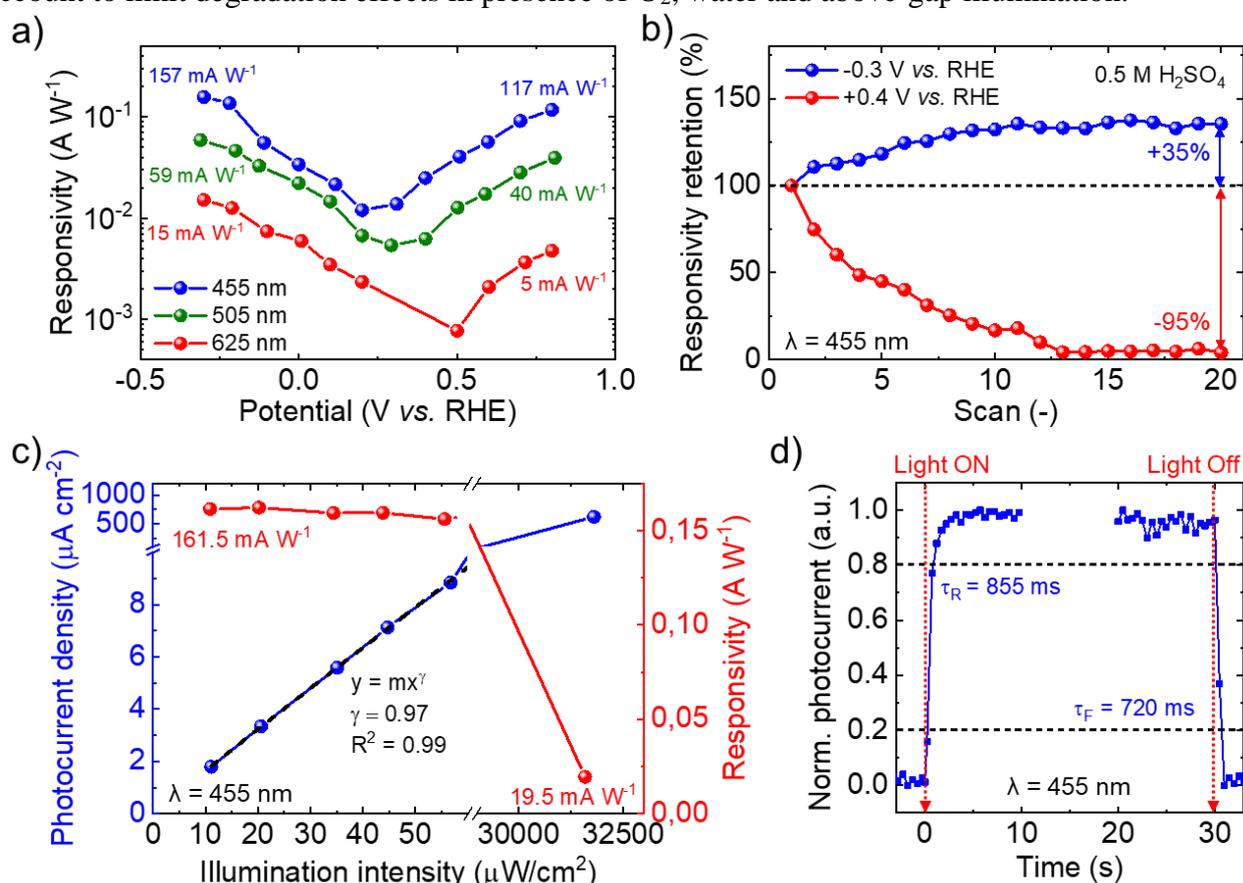

**Figure 6.** a) Responsivity of PEC-type GaSe photodetectors in 0.5 M $H_2SO_4$ as a function of the applied potential upon three different illumination wavelengths in the visible spectral range: 455, blue; 505, green; 625

nm, red. Light intensity: 63.5 µW cm$^{-2}$. b) Responsivity retention of the GaSe photodetecotors in 0.5 M H$_2$SO$_4$ at cathodic and anodic operations, *i.e.*, applied potential of –0.3 V *vs*. RHE and +0.4 V *vs*. RHE, respectively. c) Photocurrent density (left y-axis) and responsivity of the GaSe photodetectors at –0.3 V *vs*. RHE as a function of the light intensity. The curve fitting the data measured at low-light intensity is also shown (dashed black line). d) Normalized photocurrent of the GaSe photodetector at applied potential of –0.3 V *vs*. RHE measured over time after an illumination pulse of 30 s (wavelength = 405 nm, light intensity = 63.5 µW cm$^{-2}$). The rise and fall time of the photocurrent response are also indicated.

## 3. Conclusion

In summary, the electronic structure of solution processed GaSe flakes was theoretically studied using density functional theory (DFT) calculations. The predicted bandgap ($E_g$) values (3.14 eV for single-layer flake, between 2.7 and 2.0 eV for few-layer flakes) indicate that GaSe flakes can absorb a significant fraction of solar irradiation. By referring the calculated conduction band minimum energy ($E_{CBM}$) and the valence band maximum energy ($E_{VBM}$) of GaSe flakes *versus* the redox potentials of H$^+$/H$_2$ and O$_2$/H$_2$O, respectively, GaSe flakes with less than 4 layers are predicted to be pH-universal photocatalysts for hydrogen evolution reaction (HER). Meanwhile, GaSe flakes with at least 4 layers are expected as pH-universal photocatalysts for oxygen evolution reaction (OER). These results indicate that mixed nanoflakes with different number of layers could operate as "local" tandem water splitting systems. Moreover, the representative cases of bulk GaSe and trilayer-GaSe should attain the overall water splitting at pH < 1.5 and pH > 7, respectively. Driven by these expectations, GaSe single-/few-layer flakes were produced by liquid phase exfoliation (LPE) in 2-Propanol (IPA) to be used as photoelectrocatalyst in aqueous media. The GaSe photoelectrodes were fabricated *via* spray coating of the as-produced GaSe flakes dispersion onto graphite paper current collectors. They exhibit catalytic activity toward water splitting reactions, *i.e.*, HER and OER in both acidic (0.5 M H$_2$SO$_4$) and alkaline (1 M KOH) media. The GaSe photoelectrodes show the best photoelectrochemical performance in 0.5 M H$_2$SO$_4$, in which they reach a cathodic photocurrent density at 0 V *vs*. RHE ($J_{0V\ vs\ RHE}$) of –9.3 µA cm$^{-2}$, a ratiometric power-saved metric for HER ($\Phi_{saved,HER}$) of 0.09%, an anodic photocurrent density at +1.23 V *vs*. RHE ($J_{1.23V\ vs\ RHE}$) of 83.4 µA cm$^{-2}$ and a ratiometric power-saved metric for OER ($\Phi_{saved,OER}$) of 0.25%. When used as PEC-type photodetectors, GaSe photoelectrodes show a responsivity up to ~0.16 A W$^{-1}$ upon 455 nm illumination at light intensity up to 63.5 µW cm$^{-2}$ (at applied potential of –0.3 V *vs*. RHE). The stability analysis of the GaSe photodetectors evidences a durable operation in 0.5 M H$_2$SO$_4$ for PEC HER. *Viceversa*, degradation effects have been observed in both alkaline and anodic operation (*i.e.*, during OER) due to highly oxidizing environment and O$_2$ evolved-induced (photo-)oxidation effects. Our results represent the first PEC characterization of GaSe single-/few-layer flakes, and further optimization of the PEC performance can be achieved by engineering the photocathodes architecture and optimizing the electrolyte solution. Prospectively, advanced photoelectrodes could be achieved by optimizing the GaSe nanoflakes films in terms of thickness (and possible incorporation of conductive additives) and electrochemically accessible surface area/porosity, while other electrolyte solutions (both aqueous and inorganic media) could be used for carrying out other kind of PEC reactions. Overall, our research provides new insight into the exploitation of the PEC properties of GaSe nanoflakes, paving the way towards their use in photoelectrocatalysis, PEC-type photodetectors/(bio)sensors and other innovative optoelectronics devices.

**Acknowledgements**
This project has received funding from the European Union's Horizon 2020 research and innovation program under grant agreement No.785219-GrapheneCore2. We thank: Lea Pasquale (Materials Characterization Facility, Istituto Italiano di Tecnologia) for their support in XRD data acquisition/analysis; the Electron Microscopy facility – Istituto Italiano di Tecnologia for support in TEM and SEM data acquisition; and IIT Clean Room facility – Istituto Italiano di Tecnologia for the access to carry out SEM/EDS characterization. This work was supported by the project Advanced Functional Nanorobots (reg. No. CZ.02.1.01/0.0/0.0/15_003/0000444 financed by the EFRR). Z.S., D.S. and D.B. were supported by Czech Science Foundation (GACR No. 17-11456S) and by specific university research (MSMT No. 21-SVV/2019).

# Supporting Information

**Solution-processed GaSe nanoflake-based films for photoelectrochemical water splitting and photoelectrochemical-type photodetectors**

*Marilena Isabella Zappia,† Gabriele Bianca,† Sebastiano Bellani,†* Michele Serri, Leyla Najafi, Reinier Oropesa-Nuñez, Beatriz Martín-García, Daniel Bouša, David Sedmidubský, Vittorio Pellegrini, Zdeněk Sofer, Anna Cupolillo and Francesco Bonaccorso**

1. **Experimental**

**Materials**

Gallium (Ga) (99.9999 %) and selenium (Se) (99.999 %) were purchased from Strem Chemicals Inc. USA. Sulfuric acid (99.999%), KOH (≥ 85% purity, ACS reagent, pellets), were purchased from Sigma Aldrich.

**Theoretical calculations**

Electronic structure calculations were performed using density functional theory (DFT) with generalized gradient approximation (GGA, PBE96 parametrization scheme)[1] and hybrid functionals (HSE06)[2], as implemented in MedeA-VASP software package,[3] for bulk and 1-, 2-, 4- and 6-layer (denoted as B and 1L, 2L,...,6L) GaSe by inserting ~20 Å-thick vacuum region between the respective slabs. Dispersion interactions within the simple VdW+D3-zero damping approximation[4] were considered in combination with GGA-PBE96 functional. The basis set was extended up to the cut-off energy 400 eV to increase the accuracy (a 300 eV cut-off was considered to calculate the non-local exchange interaction within HSE06). The k-point mesh was constructed inside the first Brillouin zone with k-point spacing smaller than 0.2 Å$^{-1}$. A tetrahedron integration scheme was applied for electron density of states calculation.

**Synthesis and exfoliation of GaSe crystals**

GaSe crystals were synthetized by direct synthesis from Ga and Se elements.[5],[6] More in detail, an amount of granulated Ga and Se (15 g) with a nearly elemental stoichiometry of 1:1 were loaded in a quartz glass ampoule (25 mm × 150 mm), which was subsequently evacuated (pressure <5 × 10$^{-3}$ Pa) using a diffusion pump. The evacuated ampoule was then secured using an oxygen-hydrogen torch and then heated at 970 °C (*i.e.*, melting temperature of GaSe)[7] for 1 h (heating rate = 5 °C min$^{-1}$). The products were then cooled down to room temperature (cooling rate = 1 °C min$^{-1}$), obtaining the GaSe crystal. The GaSe nanoflakes were produced by liquid-phase exfoliation in 2-propanol (IPA) of the pulverized GaSe crystals, followed by sedimentation based separation (SBS) to remove unexfoliated material by ultracentrigufation. Experimentally, 50 mg of bulk crystals were added to 50 mL of anhydrous IPA and ultrasonicated in a bath sonicator (Branson® 5800 cleaner, Branson Ultrasonics) for 15 h. The resulting dispersions were ultracentrifuged at 700 g (Optima™ XE-90 with a SW32Ti rotor, Beckman Coulter) for 20 min at 15 °C in order to separate un-exfoliated bulk crystals (collected as sediment) from the exfoliated materials that remained in the supernatant. Then, the 80% of the supernatant was collected by pipetting, obtaining an exfoliated material dispersion.

**Materials characterization**

Scanning electron microscopy (SEM) analysis of the as-synthetized crystal was performed using a Helios Nanolab® 600 DualBeam microscope (FEI Company) and 10 kV and 0.2 nA as measurement conditions. The EDS spectra were acquired with a microscope combined with an X-Max detector and

INCA® system (Oxford Instruments), operating at 15kV and 0.8 nA. The samples were imaged without any metal coating or pre-treatment.

X-ray diffraction (XRD) measurements were acquired with a PANalytical Empyrean using Cu K$_\alpha$ radiation. The samples for XRD were prepared by depositing powder of GaSe crystal onto Si/SiO$_2$ substrates.

Transmission electron spectroscopy (TEM) images were acquired with a JEM 1011 (JEOL) TEM (thermionic W filament), operating at 100 kV. The morphological and statistical analysis was performed by using ImageJ software (NIH) and OriginPro 9.1 software (OriginLab), respectively. The samples for the TEM measurements were prepared by drop casting the as-prepared exfoliated material dispersions onto ultrathin C-on-holey C-coated Cu grids and rinsed with deionized water and subsequently dried overnight under vacuum.

The AFM images were acquired with a XE-100 AFM (Park System, Korea) by means of PPP-NCHR cantilevers (Nanosensors, Switzerland) having a tip diameter <10 nm. The images were collected in intermittent contact (tapping) mode on an area of 5×5 μm$^2$ (1024×1024 data points) using a drive frequency of ~330 kHz and keeping the working set point above 70% of the free oscillation amplitude. The scan rate for the acquisition of the images was 0.2 Hz. Gwyddion 2.53 software (http://gwyddion.net/) was used for processing the images and the height profiles, while the data were analysed by using OriginPro 2018 software. The latter was also used to carry out the statistical analysis on multiple AFM images for all the tested samples. The samples were prepared by drop-casting the as-prepared GaSe nanoflakes dispersion onto mica sheets (G250-1, Agar Scientific Ltd.) in N$_2$ and heating to 100°C for 15 min to dry the sample and remove adsorbates.

Raman spectroscopy measurements were performed by using a Renishaw microRaman Invia 1000 mounting a 50× objective, with an excitation wavelength of 532 nm and an incident power of 1 mW on the samples. For each sample, 50 spectra were collected. The samples were prepared by drop casting the as-prepared GaSe nanoflakes dispersion onto Au-coated Si/SiO$_2$ substrates and subsequently dried under vacuum.

Optical absorption spectroscopy (OAS) measurements were carried out on GaSe nanoflakes dispersion (diluted or concentrated at various concentrations of GaSe nanoflakes) by using a Cary Varian 5000 UV−vis spectrometer with integrating sphere.

The X-ray photoelectron spectroscopy (XPS) analysis is accomplished on a Kratos Axis UltraDLD spectrometer at a vacuum < 10$^{-8}$ mbar, using a monochromatic Al Kα source operating at 20 mA and 15 kV and collecting photoelectrons from a 300 × 700 μm$^2$ sample area. The charge compensation device was not used. Wide spectra were acquired at pass energy of 160 eV and energy step of 1 eV, while high-resolution spectra of Ga 2p, Ga 3d, Se 3d, O 1s, C 1s and Au 4f peaks were acquired at pass energy of 10 eV and energy step of 0.1 eV. The samples were prepared by drop-casting the dispersion of GaSe nanoflakes on an Au-coated Si chip in N$_2$ atmosphere while heating the substrate to 120°C. As-synthetized GaSe crystals were stacked onto conductive carbon tape and cleaved prior analysis. The samples were then transferred from air to the XPS chamber. Data analysis is carried out with CasaXPS software (version 2.3.19PR1.0). The energy scale was calibrated by setting the Au 4f$^{7/2}$ peak at 84.0 eV. For a better comparison of the as-synthetized crystal and the GaSe nanoflakes, the spectra of the GaSe crystal were calibrated by setting the binding energy of the sharp Se 3d doublet equal to the one obtained in the nanoflakes.

Ultraviolet photoelectron spectroscopy (UPS) with He I (hν = 21.2 eV) radiation was performed to estimate the Fermi energy level (E$_F$) and the valence band maximum of the materials under investigation. The experiments were conducted on the samples after the XPS analysis using the same equipment. A −9.0 V bias was applied to the sample in order to precisely determine the low kinetic energy cut-off. The energy scale was corrected according to the binding energy calibration performed for the XPS measurement.

**Electrodes fabrication**

The photoelectrodes were produced by spray-coating the GaSe nanoflakes dispersion (GaSe nanoflakes concentration = 0.13 g L$^{-1}$) onto graphite paper (PGS, Panasonic) mounted on a hot plate heated at 60 °C. The material mass loading was 0.1 mg cm$^{-2}$. The electrode area was 1.5×1 cm$^2$. The photoelectrodes were dried overnight at room temperature before the characterization.

**Electrodes characterization**

Scanning electron microscopy analysis of the of the as-produced electrodes was performed using a Helios Nanolab 600 DualBeam microscope (FEI Company) operating at 5 kV and 0.2 nA. The electrodes were imaged without any metal coating or pre-treatment.

The electrochemical measurements were performed at room temperature in a flat-bottom fused silica cell using the three-electrode configuration of the potentiostat/galvanostat station (VMP3, Biologic), controlled *via* own software. A glassy carbon rod and a saturated KCl Ag/AgCl were used as the counter-electrode and the reference electrode, respectively. The measurements were carried out in 200 mL of 0.5 M H$_2$SO$_4$ (99.999% purity, Sigma Aldrich) or 1 M KOH (99.999% purity, Sigma Aldrich). Before starting the measurements, the oxygen was purged from electrolyte by flowing N$_2$ gas throughout the liquid volume using a porous frit. A constant, slight nitrogen flow is maintained afterwards for the whole duration of the experiments, to avoid re-dissolution of molecular oxygen in the electrolyte. The Nernst equation: $E_{RHE} = E_{Ag/AgCl} + 0.059 \times pH + E^0_{Ag/AgCl}$, where $E_{RHE}$ is the converted potential *vs*. RHE, $E_{Ag/AgCl}$ is the experimental potential measured against the Ag/AgCl reference electrode, and $E^0_{Ag/AgCl}$ is the standard potential of Ag/AgCl at 25 °C (0.1976 V *vs*. RHE), was used to convert the potential difference between the working electrode and the Ag/AgCl reference electrode to the reversible hydrogen electrode (RHE) scale. A 300 W Xenon light source LS0306 (Lot Quantum Design), equipped with AM1.5G filters, was used to simulate solar illumination (1 sun). The light emitting diodes (LEDs) M455L3 (Thorlabs), M505L3 (Thorlabs) and M625L3 (Thorlabs) were used as monochromatic source. The light intensity of the LEDs was adjusted through source meter (2612B Dual-Channel System SourceMeter, Keithley)-controlled LED driver (LEDD1B, Thorlabs). The illumination intensity of the LED was calibrated by using an optical power and energy meter (PM100D, Thorlabs). The LSV curves were acquired at 5 mV s$^{-1}$ scan rate, in both anodic and cathodic directions. In agreement with ref. [8], the main Figure of Merit (FoM) used to characterize the photocathodes for PEC water splitting are: the onset potential ($V_{OP}$), defined as the potential at which the photocurrent related to the HER (hydrogen evolution reaction) or OER (oxygen evolution reaction) is observed; the cathodic photocurrent density at 0 V *vs*. RHE ($J_{0V\,vs\,RHE}$); the anodic photocurrent density at 1.23 V *vs*. RHE ($J_{0V\,vs\,RHE}$), the ratiometric power-saved metric for HER ($\Phi_{saved,HER}$) and the ratiometric power-saved metric for OER ($\Phi_{saved,OER}$). The ratiometric power-saved metrics are calculated by $\Phi_{saved} = \eta_F \times |j_{photo,m}| \times [E_{light}(J_{photo,m}) - E_{dark}(J_{photo,m})]/P_{in} = \eta_F \times |J_{photo,m}| \times V_{photo,m}/P_{in}$, in which $\eta_F$ is the current-to-hydrogen (or oxygen) faradaic efficiency assumed to be 100%, $P_{in}$ is the power of the incident illumination, and $j_{photo,m}$ and $V_{photo,m}$ are the photocurrent and photovoltage at the maximum power point, respectively. $j_{photo}$ is obtained by calculating the difference between the current under illumination of a photocathode and the current of the corresponding catalyst. The photovoltage $V_{photo}$ is the difference between the potential applied to the photocathode under illumination ($E_{light}$) and the potential applied to the catalyst electrode ($E_{dark}$) to obtain the same current density. The subscript "m" stands for "maximum".

The stability of the GaSe photodetectors was evaluated by recording subsequent 20 LSV scans and measuring the corresponding responsivity.

## 2. Layer-dependent bandgap of GaSe

**Figure S1** shows the GaSe bandgap ($E_g$) as a function of the number of layers (plot derived by electronic structure calculations performed using DFT with hybrid functionals (HSE06 functional approach)[2] for bulk and 1-, 2-, 4- and 6-layer (denoted as B and 1L, 2L,..,6L)-GaSe (see details in

Section 1, Experimental). By reducing the number of layers in the GaSe, the $E_g$ progressively increases up to the highest value in 1L-GaSe (3.14 eV).

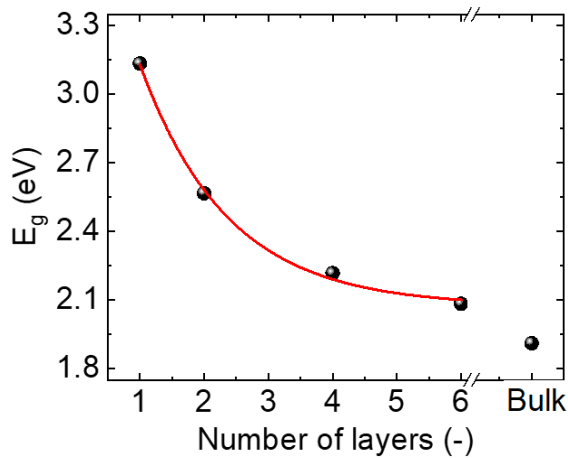

**Figure S1.** Plot of the theoretically calculated GaSe $E_g$ versus the number of layer. The $E_g$ calculated for the B-GaSe is also included.

3. **Scanning electron microscopy-coupled energy dispersive X-ray spectroscopy analysis of as-synthetized GaSe crystals**

**Table S1** reports the chemical composition of the as-synthetized GaSe crystals, as derived from the scanning electron microscopy (SEM)-coupled energy dispersive X-ray spectroscopy (EDS) analysis reported in the main text (**Figure 3b,c**).

As discussed in the main text, the SEM-EDS analysis reveals a slight Ga-enriched phases of the GaSe crystals (Ga-to-Se atomic ratio ~1.4), which is in agreement with previous studies.[5],[6] The stoichiometric excess of Ga is attributed to the formation oxides (*i.e.*, $Ga_2O_3$), which partially passivate the GaSe surface and prevent the underlying GaSe from further oxidation. [9],[5],[10]

**Table S1.** Elemental composition of the as-synthetized GaSe crystals obtained from SEM-coupled EDS analysis.

| Element | atomic % |
|---------|----------|
| Ga      | 59.1     |
| Se      | 40.9     |

## 4. X-Ray diffraction measurement of the as-synthetized GaSe crystals

**Figure S2** shows that XRD pattern of the GaSe crystals corresponds to the JCPDS 37-931 card. This means that the as-synthetized crystals are in form of the lowest energy polytypes, *i.e.*, the hexagonal $\varepsilon$-GaSe[11],[12] (space symmetry group: $P\bar{6}m$-$D'_{3h}$),[13],[14],[15] in agreement with previous literature reporting similar GaSe crystal syntheses.[5],[6]

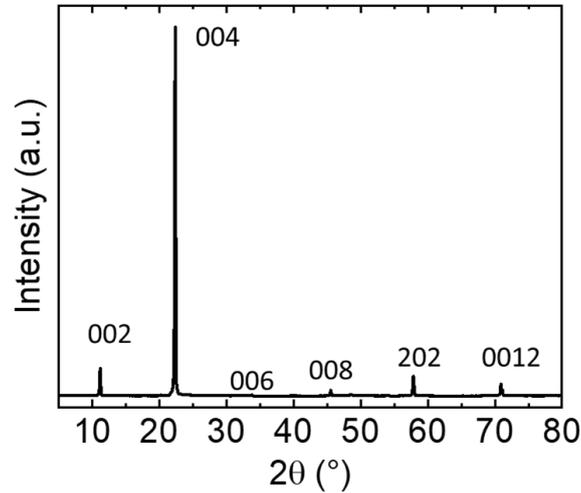

**Figure S2.** XRD pattern of the as-synthetized GaSe crystals. The XRD peaks assigned to $\varepsilon$-GaSe are also shown according to the JCPDS 37-931 card.

## 5. Raman statistical analysis

**Figure S3** reports the Raman statistical analysis of the $A^1_{1g}$ and $E^1_{2g}$ Raman peaks of the as-synthetized GaSe crystals (panels a,b) and GaSe nanoflakes (panel c,d). In agreement with both theoretical[16] and experimental[16],[17],[18] studies on the thickness dependence of Raman spectrum of GaSe, $A^1_{1g}$ and $E^1_{2g}$ slightly shifts at lower and higher wavenumbers, respectively, passing form the GaSe crystal to the GaSe nanoflakes.

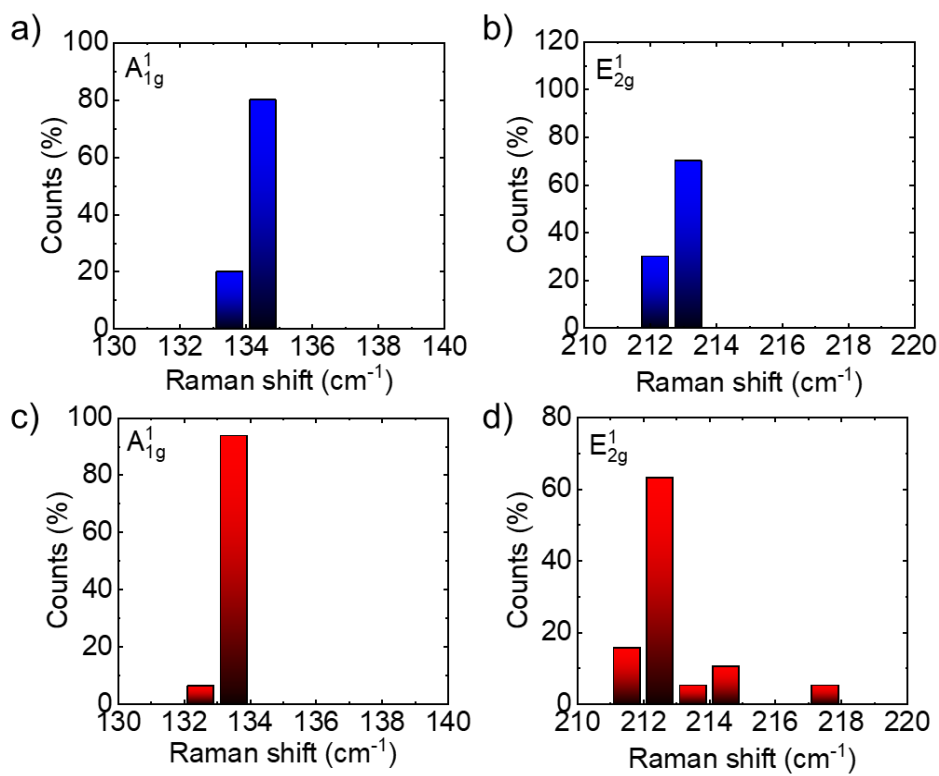

**Figure S3.** Raman statistical analysis of the $A^1_{1g}$ and $E^1_{2g}$ peak for a,b) the as-synthetized GaSe crystals and c,d) the GaSe nanoflakes.

## 6. X-ray photoelectron spectroscopy analysis of the as-synthetized GaSe crystals and the GaSe nanoflakes

### 6.1 X-ray photoelectron spectroscopy characterization of the as-synthetized GaSe crystal

The wide XPS spectrum of the as-synthetized GaSe crystals (**Figure S4a**) shows peaks related to Ga, Se, C and O, attributable to the GaSe material, oxidized species and contamination from conductive carbon tape or adventitious hydrocarbons. The results of the elemental analysis are shown in **Table S2**. Ga and Se are present in a ratio of 59.7:40.3, resulting in a stoichiometry $GaSe_{0.68}$, in agreement with the SEM-EDS analysis. The deviation from the ideal stoichiometry may be due to surface oxidation of the crystal leading to the formation of an oxide layer and the displacement of Se, or to the presence of Se vacancies in the GaSe structure. High resolution spectra of the Ga 2p, Ga 3d, Se 3d, O 1s, C 1s regions are shown in **Figure S4b–f**. The deconvolution of the Ga 2p and 3d peaks (**Figure S4b,c**) leads to the identification of three chemical states of Ga, attributable to GaSe (binding energy –B.E.–: ~1117.6 eV for $2p^{3/2}$; ~19.5 eV for $3d^{5/2}$), Ga(+III) (*i.e.*, gallium oxides and $Ga_2Se_3$ oxides (B.E.: ~1118.3 eV for $2p^{3/2}$; ~20.5 eV for 3d)[19–21] and metallic Ga (Ga(0)) (B.E.: ~1116.5 eV for $2p^{3/2}$; 18.5 eV for 3d).[19,20] The value of spin-orbit splitting ($\Delta_{so}$) found for the 2p doublet is 26.87 eV, while for the unresolved 3d doublet the best fit value of $\Delta_{so}$ is 0.51 eV, in agreement with the literature.[20,22] Metallic gallium could form by disproportionation of GaSe toward oxidized species or originated by residuals of synthesis precursors. The discrepancy between the amounts of Ga species obtained from the Ga 2p and Ga 3d may be due to an inhomogeneous distribution of these species and to the different sampling depth of these XPS regions. The Ga 2p photoelectrons have lower kinetic energy (K.E.) (~369 eV) and shorter inelastic mean free path ($\lambda$) (~1.1 nm)[23,24] than the Ga 3d photoelectrons (K.E. ~1466 eV, $\lambda$ ~3.0 nm), hence the latter probes deeper layers below the surface compared to Ga 2p. As expected, the amount of Ga(+III) is 39.8% for the Ga 2p and decreases to 20.4% for the Ga 3d, confirming that the oxidation is confined to the surface. The distribution of Ga(0), detected in smaller amounts (**Table S3**), follows a similar trend.

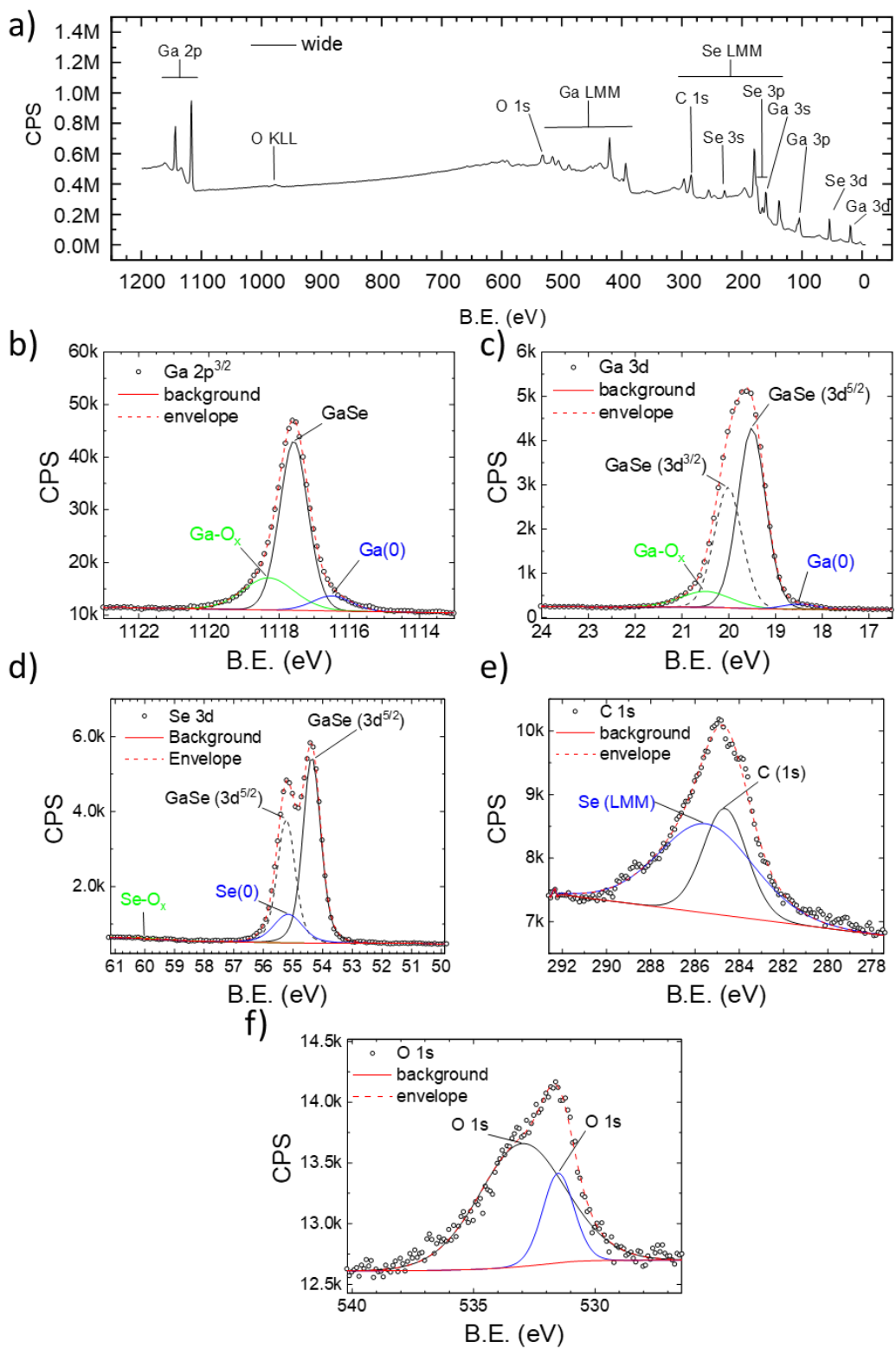

**Figure S4.** Wide XPS spectra of the as-synthetized GaSe crystal a) and high resolution spectra of the Ga 2p b), Ga 3d c), Se 3d d), C 1s e) and O 1s f) regions with their deconvolution fit.

The Se 3d peaks can be deconvoluted into three components, attributable to GaSe (B.E. ~54.4 eV $3d^{5/2}$, $\Delta_{so} = 0.86$ eV), Se(0) (B.E. ~55.2 for 3d)[22], and selenium oxides (Se-Ox) (B.E. ~59.9 eV for 3d)[22] (**Figure S4d**). The relative amounts, shown in **Table S4**, indicate that only a modest quantity of Se(0) (~13%) is present on the surface as a product of GaSe oxidation, while Se-Ox species are virtually absent. The C 1s region (**Figure S4e**) shows overlap between the carbon signal and Auger peaks of selenium. The deconvolution returns a C 1s peak that was set to B.E. ~284.8eV. The oxygen O 1s signal shows two component (**Figure S4f**), compatible with functional groups of contaminants and the aforementioned presence of surface Ga(+III).

**Table S2.** XPS elemental analysis of the as-synthetized GaSe crystal

|  | Ga (3d) | Se | C | O | Total |
|---|---|---|---|---|---|
| % (Ga, Se, C, O) | 33.2 | 22.4 | 30.7 | 13.8 | 100 |
| % (Ga, Se) | 59.7 | 40.3 |  |  | 100 |

**Table S3.** Relative amount of different chemical states of gallium in the as-synthetized GaSe crystal determined from deconvolution of Ga 2p and Ga 3d peaks.

|  | GaSe | Ga(+III) | Ga(0) | Total |
|---|---|---|---|---|
| %(Ga 2p) | 68.4 | 23.3 | 8.3 | 100 |
| %(Ga 3d) | 89.6 | 8.6 | 1.8 | 100 |

**Table S4.** Relative amount of different chemical states of selenium in the as-synthetized GaSe crystal determined from deconvolution of Se 3d peak.

|  | GaSe | Se-Ox | Se(0) | Total |
|---|---|---|---|---|
| %(Se 3d) | 86.4 | 0.6 | 13.0 | 100 |

## 6.2 XPS characterization of the exfoliated GaSe nanoflakes

The wide XPS spectrum of the GaSe nanoflakes (**Figure S5a**) shows peaks related to Ga, Se, C, O and Au, attributable to the GaSe material, oxidized species, carbon contamination from solvent residues or adventitious hydrocarbons as well as the substrate. The results of the elemental analysis are shown in **Table S5**. Ga and Se are present in a ratio of 58.7:41.3, resulting in a stoichiometry GaSe$_{0.7}$, similar to the as-synthetized crystal. The deviation from the ideal stoichiometry may be due to surface oxidation of the nanoflakes leading to the formation of a Ga oxide layer and the displacement of Se, or to the presence of Se vacancies in the GaSe structure. High resolution spectra of the Ga 2p, Ga 3d, Se 3d, O 1s, C 1s regions are shown in **Figure S5b–f**. Similar to the GaSe crystal, the deconvolution of the Ga 2p and 3d peaks (**Figure S5b,c**) leads to the identification of three chemical states of Ga, attributable to GaSe (B.E.: ~1117.6 eV for 2p$^{3/2}$; ~19.5 eV for 3d$^{5/2}$), Ga(+3) (B.E.: ~1118.3 eV for 2p$^{3/2}$; ~20.5 eV for 3d)[19,20] and Ga(0) (B.E.: ~1116.5 eV for 2p$^{3/2}$; ~18.5 eV for 3d).[19,20] The value of $\Delta_{so}$ for the 2p and 3d doublets are the same as in the GaSe crystals. As expected, the GaSe nanoflakes present a higher content of Ga(+III) (**Table S6**) compared to the GaSe crystals, due to a higher surface/volume ratio. The depth distribution of Ga species inferred from the comparison of Ga 2p and Ga 3d is similar to the one observed in the crystal, suggesting that the GaSe nanoflakes are oxidized only on the surface.

The Se 3d peaks can be deconvoluted in three components, attributable to GaSe (B.E. ~54.4 for 3d$^{5/2}$, $\Delta_{so}$ = 0.86 eV), Se (0) (B.E. ~55.23 for d), and Se-Ox (B.E. ~59.9 eV for 3d) (**Figure S5d**). The relative amounts of Se species are shown in **Table S7** and indicate that only a modest quantity of Se(0) ~11.1% is present on the surface as a product of GaSe oxidation, while Se-Ox species are absent. The C 1s region (**Figure S5e**) shows overlap between the carbon signal and Auger peaks of Se. The deconvolution returns a C 1s B.E. ~284.3eV, slightly lower than the typical value of hydrocarbons (284.6-285 eV) possibly due to inaccuracy of the deconvolution model or small charging effects. The oxygen O 1s signal shows a single component (**Figure S5f**), compatible with the aforementioned presence of surface Ga(+3).

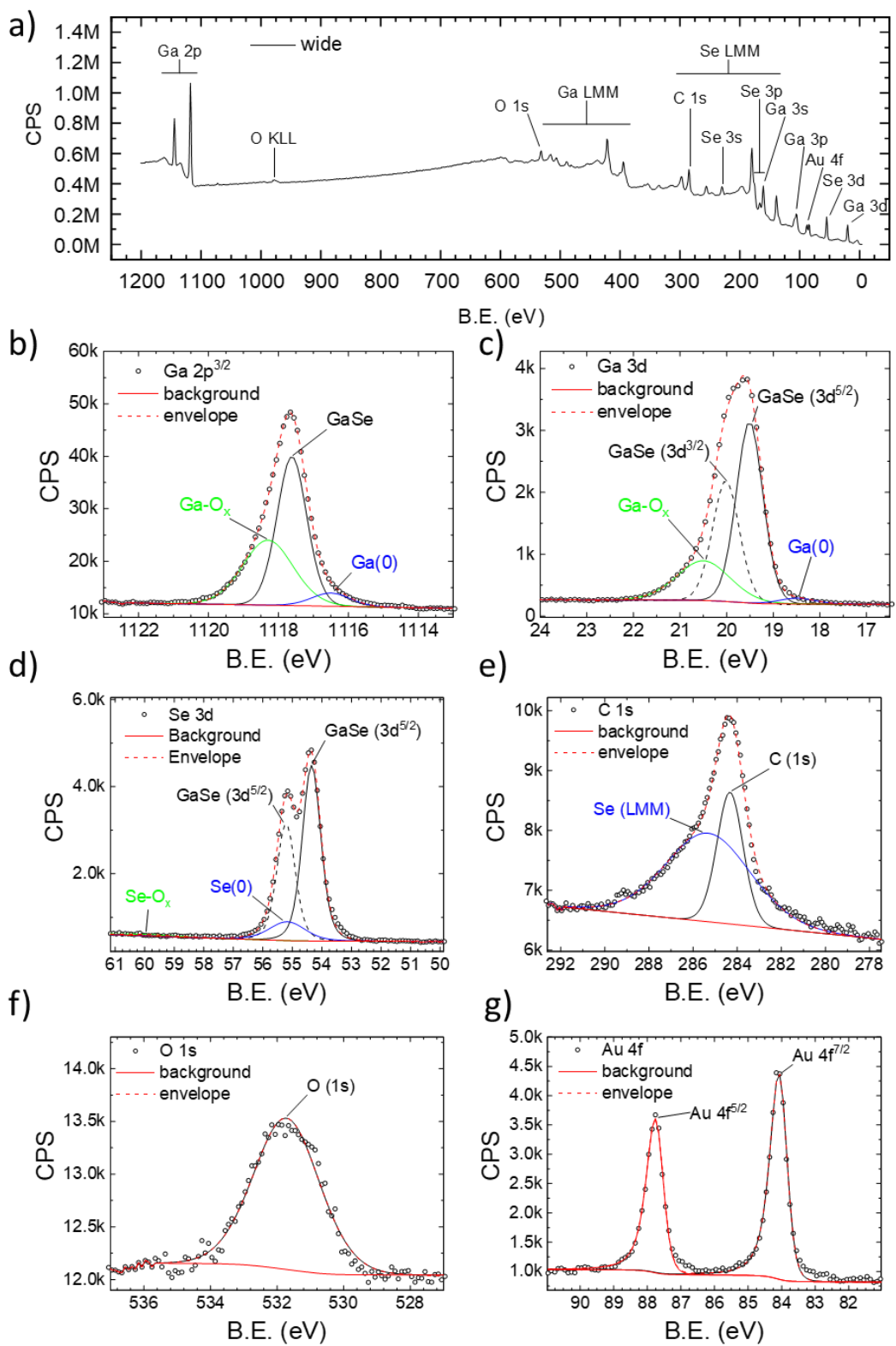

**Figure S5.** Wide XPS spectra of GaSe nanoflakes a) and high-resolution spectra of the Ga 2p b), Ga 3d c), Se 3d d), C 1s e), O 1s f) and Au 4f g) regions with their deconvolution fit.

**Table S5.** XPS elemental analysis of GaSe nanoflakes

| | **Ga (3d)** | **Se** | **C** | **O** | **Total** |
|---|---|---|---|---|---|
| % (Ga, Se, C, O) | 34.1 | 24.0 | 30.8 | 11.1 | 100 |
| % (Ga, Se) | 58.7 | 41.3 |  |  | 100 |

**Table S6.** Relative amount of different chemical states of gallium in GaSe nanoflakes determined from deconvolution of Ga 2p and Ga 3d peaks.

|  | GaSe | Ga(+III) | Ga(0) | Total |
| --- | --- | --- | --- | --- |
| %(Ga 2p) | 54.0 | 39.8 | 6.2 | 100 |
| %(Ga 3d) | 77.6 | 20.4 | 2.0 | 100 |

**Table S7.** Relative amount of different chemical states of selenium in GaSe nanoflakes determined from deconvolution of Se 3d peak.

|  | GaSe | Se-Ox | Se(0) | Total |
| --- | --- | --- | --- | --- |
| %(Se 3d) | 87.7 | 1.3 | 11.1 | 100 |

## 7. Comparison between the responsivity of our EC-type GaSe photodetectors and other relevant solution-processed photodetectors reported in literature.

**Table S8.** Comparison between the responsivity our PEC-type GaSe photodetectors with the one of other solution-processed photodetectors reported in literature.

| Materials | Device configuration | Measurement conditions | | Responsivity (mA W$^{-1}$) | Illumination intensity (mW cm$^{-2}$) | Wavelength (nm) | Reference |
|---|---|---|---|---|---|---|---|
| | | Electrolyte | Applied potential | | | | |
| GaSe nanoflakes | PEC-type | 0.5 M H$_2$SO$_4$ | -0.3 V vs. Ag/AgCl | ~160 | < 0.0567 | 455 | This work |
| | | | | 19.5 | 31.8 | 455 | |
| InSe nanosheets | PEC-type | 0.2 M KOH | 1 V vs. SCE | 3.3 × 10$^{-3}$ | 120 | Simulated sunlight | [25] |
| | | | | 4.9 × 10$^{-3}$ | 40 | Simulated sunlight | |
| Black phosphorous nanosheets | PEC-type | 0.1 M KOH | 0 V vs. SCE | 1.9 × 10$^{-3}$ | 20 | Simulated sunlight | [26] |
| | | | | 2.2 × 10$^{-3}$ | 100 | Simulated sunlight | |
| GeSe nanosheets | PEC-type | 0.1 M KOH | 0.3 V | 0.044 | 118 | Simulated sunlight | [27] |
| | | | | 0.076 | 26.2 | Simulated sunlight | |
| SnS | PEC-type | 0.1 Na$_2$SO$_4$ | 0.6 V | 0.018 | 3.57 | 365 | [28] |
| Perovskite (CH$_3$NH$_3$PbI$_3$) | Metal-semiconductor-metal | - | 5 V | 4.4 | 1 | 633 | [29] |
| PBDTT-ffQx/PCBM bulk heterojunction | Metal-semiconductor-metal | - | 10 V | 1.15 × 10$^3$ | 25 | 365 | [30] |

| SnS/RGO hybrid nanosheets | FET | - | $V_{DS}$ =5V, $V_g$ = 0 V | 180 | 0.12 | Visible light | [31] |
|---|---|---|---|---|---|---|---|
| Perovskite (CH$_3$NH$_3$PbI$_3$) PDPP3T | Metal-semiconductor-metal | - | 1 V | 10.7 | 0.5 | 365 | [32] |
| | | | | 25.5 | | 650 | |
| | | | | 5.5 | | 937 | |

## 8. Characterization of photoelectrochemical-type GaSe photodetector in 1 M KOH

**Figure S6a** shows the responsivity of the GaSe photodetector in 1 M KOH as a function of the applied potential at wavelengths of 455 nm (blue) and 625 nm (red) (light intensity = 63.5 µW cm$^{-2}$). Since the stability issues observed in cathodic regime for HER (see **Figure 5a** in the main text), only positive applied potentials were investigated in 1 M KOH. As expected from the OAS analysis (**Figure 3i** in the main text) and similarly to the photodetectors in 0.5 M, the responsivity of the GaSe photodetectors shows the maximum value (41.9 mA W$^{-1}$ at +1.3 V *vs.* RHE) under illumination at 455 nm. In agreement with the PEC water splitting activity under simulated sunlight (see **Figure 5b** in the main text), the responsivity of the GaSe photodetectors in 1 M KOH is lower than the one measured in 1 M H$_2$SO$_4$. Moreover, as shown in **Figure S6b**, the GaSe photodetector progressively degrades over different LSV scans (–97% after 20 LSV scans). As for the GaSe photodetector in 1 M H$_2$SO$_4$, the degradation under anodic potentials could be originated by the progressive oxidation of the GaSe nanoflakes caused by both anodic potential and evolved O$_2$.[5]

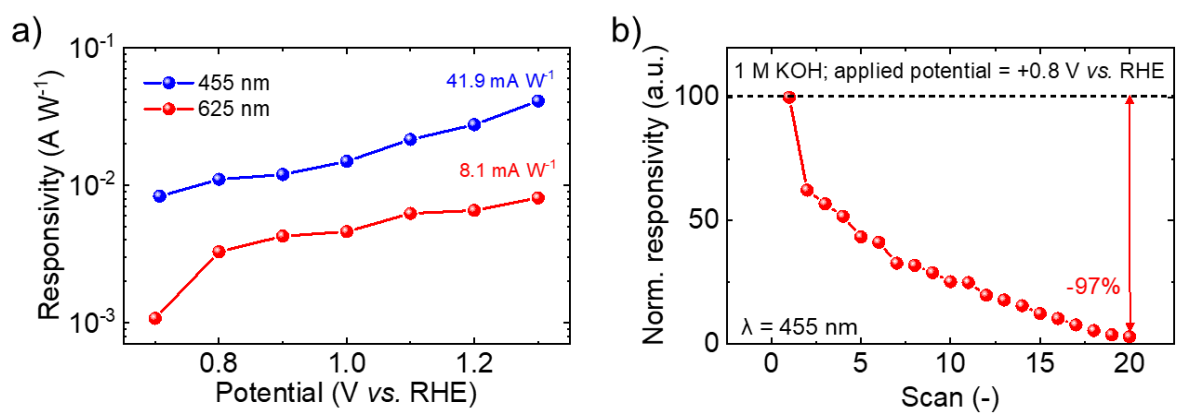

**Figure S6.** a) Responsivity of PEC-type GaSe photodetectors in 1 M KOH as a function of the applied potential upon illumination wavelengths of 455, blue and 625 nm, red. Light intensity: 63.5 µW cm$^{-2}$. b) Responsivity retention of the GaSe photodetectors in 1 M KOH at applied potential of +0.8 V *vs.* RHE, respectively.

## 9. Raman spectroscopy measurement of the GaSe electrodes before and after PEC stability tests

**Figure S7** show the Raman spectra of GaSe photoelectrode before and after the stability tests shown in **Figure 6b** and **Figure S6**. The fresh electrode (**Figure S7a**) shows the peaks attributed to $Ga_2O_3$ and elemental Se, which originated by the air-induced oxidation of GaSe nanoflakes during their spray coating deposition onto the electrode current collector. As discussed in the Intorduction section of the main text, the oxidation process can occur in a two-step reaction: $GaSe + 1/4O_2 = 1/3Ga_2Se_3 + 1/6Ga_2O_3$ followed by $Ga_2Se_3 + 3/2O_2 = Ga_2O_3 + 3Se$;[33] or in a single-step reaction: (2) $GaSe + 3/4O_2 = 1/2Ga_2O_3 + Se$.[34] However, the material oxidation process can be controlled by properly adjusting (photo)electrochemical conditions such as potential, pH and the dissolved $O_2$ concentration.[35] In particular for our case, the Raman spectrum of the GaSe photoelectrodes after stability test for HER in 0.5 M $H_2SO_4$ show the peak attributed to GaSe, indicating that GaSe nanoflakes preserve their structural properties, while the peak ascribed to $Ga_2O_3$ disappears. In agreement with the Pourbaix diagram of Ga,[36] the disappearance of $Ga_2O_3$ peak can be explained by the corrosion of $Ga_2O_3$, which dissolves into $Ga^{2+}$. Consequently, the electrode self-optimizes the exposure of the electrocatalytically active GaSe nanoflakes, showing an increase of their HER-activity during successive cathodic LSV scans (see **Figure 6b** in the main text). *Viceversa*, after anodic operation, the electrode still shows the peak attributed to $Ga_2O_3$. The presence of $Ga_2O_3$, as well as the decrease of the OER-activity observed during stability tests, indicate that GaSe nanoflakes are susceptible to a chemical degradation during anodic operation. After stability test for OER in 1 M KOH, the peak attributed to $Ga_2O_5$ is not observed, since $Ga_2O_5$ can be corroded to form $GaO_2^-$ and/or $GaO_3^{3-}$.[36] However, the high intensity of the peak attributed to elemental Se, as well as the OER-activity degradation (**Figure S6**), support that GaSe nanoflakes can decompose during anodic operation. As discussed in the the main text, the photoelectrode degradation during OER (in both acidic and alkaline media) can be reasonably attributed to the progressive oxidation of the GaSe flakes caused by both anodic potentials and evolved $O_2$.

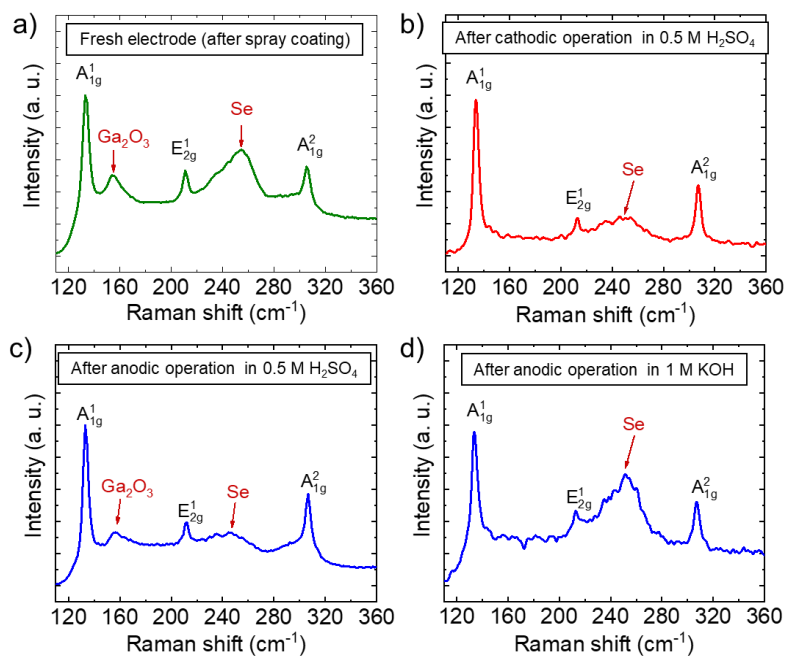

**Figure S7.** Raman spectra of GaSe electrodes a) before (as-fabricated GaSe electrodes) and b-d) after stability tests for: HER in 0.5 M H2SO4 (panel b); OER in 0.5 M H2SO4 (panel c) and OER in 1 M KOH (panel d). The Raman peaks attributed to GaSe, Ga2O3 and elemental Se are indicated.